\documentclass[preprint2]{aastex631}

\DeclareUnicodeCharacter{2060}{}

\usepackage{lmodern}
\usepackage{anyfontsize}
\usepackage{amsmath}
\usepackage{ragged2e} 

\begin{document}

\title[Short title, max. 45 characters]{Identifying a protocluster formation at $z\sim$ 4.5 in the COSMOS field: an extension of Taralay protocluster, 
traced by Ly$\alpha$ emitters surrounding a submm galaxy}

\correspondingauthor{Mariana Rubet}
\email{marianarubet@ov.ufrj.br}

\author[0000-0001-9439-3121]{Mariana Rubet}
\affiliation{Observatorio do Valongo, Universidade Federal do Rio de Janeiro \\
Ladeira Pedro Antonio 43, Rio de Janeiro, RJ 20080-090, Brazil}

\author[0000-0003-3153-5123]{Karín Menéndez-Delmestre}
\affiliation{Observatorio do Valongo, Universidade Federal do Rio de Janeiro \\
Ladeira Pedro Antonio 43, Rio de Janeiro, RJ 20080-090, Brazil}

\author[0000-0003-2374-366X]{T.S. Gonçalves}
\affiliation{Observatorio do Valongo, Universidade Federal do Rio de Janeiro \\
Ladeira Pedro Antonio 43, Rio de Janeiro, RJ 20080-090, Brazil}

\author[0000-0002-8048-8717]{F. Almeida-Fernandes}
\affiliation{Observatorio do Valongo, Universidade Federal do Rio de Janeiro \\
Ladeira Pedro Antonio 43, Rio de Janeiro, RJ 20080-090, Brazil}
\affiliation{Instituto Nacional de Pesquisas Espaciais, \\ Av. dos Astronautas 1758, Jardim da Granja,12227-010 S\~ao Jos\'e dos Campos, SP, Brazil}

\author[0000-0003-3402-6164]{L. Santana-Silva}
\affiliation{Observatorio do Valongo, Universidade Federal do Rio de Janeiro \\
Ladeira Pedro Antonio 43, Rio de Janeiro, RJ 20080-090, Brazil}

\author[0000-0003-3578-6843]{Peter L. Capak}
\affiliation{Cosmic Dawn Center (DAWN), Denmark}

\author[0000-0002-5496-4118]{Kartik Sheth}
\affiliation{NASA Headquarters, 300 Hidden Figures Way SW, Washington DC 20546, USA}

\begin{abstract}

We report the discovery of a large-scale structure containing multiple overdensities at $z= 4.54 \pm 0.03$ in the COSMOS field, with the most prominent one likely infalling towards the recently identified Taralay protocluster. We use combined wide-band and narrow-band optical photometry to identify Ly$\alpha$ emitters (LAEs) within a 21 cMpc radius from the submm source J1000+0234, at $z= 4.54$, to identify typical star-forming galaxies that may trace an underlying structure. Our approach selects line emitters as narrow-band excess objects and we use the COSMOS2020 photometric redshift catalog to eliminate potential low-redshift interlopers whose line emission (e.g. [OIII] at $z\sim 0.3$) might be responsible for the observed excess in the narrow band. In comparison with the LAE density in the field, our results point to a mean LAE number overdensity of $\bar\delta = 3$ spanning a region of $27 \times 20 \times 36$ cMpc$^3$, probably evolving into a moderate-mass cluster ($3$–$10 \times 10^{14} \, M_\odot$) at $z\sim 0$. This work supports the idea that submm sources, although offset from the major overdensity peaks, serve as traces of moderately massive, potentially infalling structures.

\end{abstract}


\section{Introduction}
\label{sec:intro}

Initial fluctuations in the primordial density field, amplified by anisotropic gravitational collapse driven by dark matter, initiated the process of large-scale structure formation (see, e.g.,  \citealp{Springel_2006}), progressively giving rise to the observed cosmic web patterns such as filaments, nodes, sheets and voids. The densest structures in the local Universe, galaxy clusters, form at the nodes of the cosmic web, through a hierarchical sequence of mergers and accretion of smaller systems \citep[see e.g.,][]{Boylan-Kolchin_2009}. They reach masses greater than $10^{14}$ M$_{\odot}$ and consist of a virialized dark matter halo, comprising a hot, X-ray-emitting intracluster medium, which surrounds hundreds to thousands of galaxies - including red, passive systems \citep{Kravtsov_2012,2015MNRAS.452.2528M}.

The relevance of environmental effects on galaxies has been reported across a wide range of environments, from the densest ones, such as clusters in the local Universe (\citealp[e.g.,][]{1980ApJ,Dressler_1997,Postman_2005}), to less dense regions such as filaments and groups \citep[e.g.,][]{Postman_&_Geller1984,Smith_2005}, in which significant environmental processing is reported to occur on galaxies prior to cluster infall \citep[e.g.,][]{Castignani_2022,Lopes_2023, MAGIC_2024}. The early, non-virialized, and sparse stages of clusters are known as protoclusters. These structures provide a valuable setting to investigate how environmental effects progressively emerge in overdense regions \citep[e.g.,][]{Ryley_2022}, as well as the physical mechanisms responsible for shaping the observed galaxy properties \citep[e.g.,][]{Fujita_2004,Boselli_2006}.

The search for protoclusters of galaxies is challenging because these are sparse structures at high redshifts extending over tens of Mpc \citep{2015MNRAS.452.2528M} and hosting weakly bound galaxies. These structures often lack an intracluster hot gas or a well-formed red sequence, that can otherwise be seen in evolved clusters \citep{Gladders_2000, Rosati_2002, Andreon_2009, Mantz_2014}. Additionally, due to the gradual process of cluster assembly, the density contrast between protoclusters and the field is relatively small at high redshift, requiring highly sensitive surveys and, preferably, spectroscopic coverage to confirm whether galaxies truly belong to the structure or if they are low-redshift interlopers mistaken for members due to projection effects. One strategy to identify these highly sparse structures is to search for objects with a high probability of inhabiting massive halos at high redshift. The initial protocluster searches were conducted using radio galaxies, which, being hosted by high-mass galaxies ($M_{*}>10^{11}M\odot$) at $z=1-4$ \citep{Seymour_2007}, are expected to inhabit the most massive halos. 
 Several protoclusters were discovered using this approach (e.g., \citealp{Le_Fevre}; \citealp{Venemans_2006}; \citealp{Cooke_2014}). 
However, due to the relatively short timescale during which we can observe the activity of a radio galaxy (\citealp[i.e., $\sim10^7$ years, e.g.,][]{Blundell_1999}), 
many protoclusters are likely missed by this method. This highlights the need for other tracers. Another class of objects used as targets are QSOs; however, different studies indicate that they inhabit both overdense and more typical regions (e.g., \citealt{Hennawi_2015}; \citealt{Uchiyama_2017}; \citealt{2023arXiv230410437C}). Ly$\alpha$ Blobs (LABs) have also helped to unveil some protoclusters. LABs are extensive luminous sources ($>50$kpc, $L_{Ly\alpha} \sim 10^{43-44} $ erg.s$^{-1}$, \citealt{2020ARA&A..58..617O}) emitting Ly$\alpha$ line radiation and are often associated with overdense regions or cosmic web filaments (\citealt{Yang2010}, \citealt{Umehata_2019}, \citealp{Ramakrishnan2023,Vandana_2024}).

We use submillimeter (submm) galaxies (SMGs) \citep[e.g.,][]{Smail_1997,Blain_1999,Blain_2002} as targets for candidate protoclusters \citep[e.g.,][]{Smol_2017,Crespo_2021}. These objects are intensely star-forming and heavily dust-obscured galaxies at high redshifts (typically $z\sim$ 1–4), with their high infrared ($>10^{12}$L$_{\odot}$) emission detected in the submillimeter portion of the spectrum due to cosmological reddening. 
Due to their high star formation rates, reaching $\sim$10$^3$M$_{\odot}$ yr$^{-1}$ \citep{Magnelli_2012A&A}, and massive gas reservoirs ($\sim$10$^{10}$M$_{\odot}$; \citealp{Greve_2005}), they are expected to inhabit density peaks and potentially be the progenitors of massive elliptical galaxies (\citealt{Lilly_1999}; \citealt{Swinbank_2006}; 
\citealt{Toft_2014};
\citealt{Gomez-Guijarro_2018};
\citealt{Dudzevi_2020}), thus making them possible tracers of protoclusters. Searches with SMGs have revealed numerous cases of SMGs residing in massive halos and protoclusters at high redshift (e.g. \citealp{Hickox_2012};
\citealp{2023arXiv230408511M};  \citealp{Calvi+23}). Other studies report SMGs in filamentary overdense environments (\citealp[e.g.,][]{sun+23}
) or inhabiting less dense regions undergoing active star formation episodes (\citealp[e.g.,][]{2009ApJ...691..560C}), and even located in sparsely populated regions, as voids in large-scale structure \citep{10.1093/mnras/stv1618}. Recently, \cite{Cornish_2024} examined the environments of three spectroscopically confirmed SMGs at $z\sim$ 2.3 and $z\sim$ 3.3. They reported that these galaxies are located in a range of environments, spanning from overdense protoclusters or protogroups to more typical field regions, suggesting that while some SMGs may evolve into massive elliptical galaxies within clusters, they may also follow alternative evolutionary paths.

Protoclusters have also been reported to be associated with overdensities of Ly$\alpha$ emitting galaxies (LAEs) and many studies have used LAEs 
to probe and quantify the overdensity of these 
regions (\citealp[e.g.,][]{Venemans_2007, Kyoung-Soo_2014, Badescu_2017,Hu_2021,Ramakrishnan2023,Vandana_2024}).
These are typically young galaxies ($\sim$10$^7$ years), with low stellar mass ($\sim$10$^{8-9}$ M$_{\odot}$), and high star formation rates of $\sim$1-2 M$_{\odot}$ yr$^{-1}$\citep[see the review][]{2020ARA&A..58..617O}. They are also less dusty than any other known galaxy population that resides in moderate-mass halos, showing low bias at high redshift (e.g., $b \sim$ 1.7,  $z\sim$ 3, \citealp{Gawiser_2007}). For these reasons, LAEs have been widely used as visible tracers to map large-scale structures (e.g., \citealp{2012_Uchimoto};  
\citealp{Kyoung-Soo_2014};
\citealp{Badescu_2017};
\citealp{2022_Huang}; \citealp{Ramakrishnan2023,Vandana_2024}).

In this work, we investigate the environment of the submm source J1000+0234 \citep{Schinnerer_2008,Scott_2008,Capak_2008} (also known as AzTEC/C17, \citealp{Aretxaga_2011_submm}) at $z=4.54$ in the Cosmic Evolution Survey field (\citealp{COSMOS_survey}) - a field largely used for probing the formation and evolution of galaxies as a function of both redshift and the galaxy environment. This field was targeted due to mounting evidence of a rich presence of SMGs. Several studies have reported an excess of high-redshift dusty star-forming galaxies in this area, including \cite{Smolcic_2015}, who analyzed the properties of six SMGs at $z>4$, among them, J1000+0234. Groups of SMGs have been identified at  $z\sim1-3$ in this field (e.g., \citealp{Aravena_2010}). Additionally, \cite{Capak_2011} identified a protocluster at $z=5.3$, surrounding the submm source AZTEC3,
which was further characterized by \cite{Riechers_2014}. Recently, \cite{Jimenez-Andrade_2023} conducted observations with the Multi-Unit Spectroscopic Explorer (MUSE) within a 48 × 48 arcsec² area around J1000+0234 and reported a galaxy overdensity that might evolve into a galaxy cluster by $z=0$. Given this evidence, we probe a larger 0.12 deg$^2$ area centered on J1000+0234 to determine whether it resides in an overdense protocluster region or a more typical, less dense area of the cosmic web. Our method uses narrow-band photometry to identify LAE candidates in a small redshift range centered on J1000+0234 ($z= 4.54 \pm 0.03$). We calculate the LAE overdensity in the surrounding region - 
using as a reference the mean LAE density in the field estimated from \cite{2020ARA&A..58..617O} - and the mass of the structure it will probably evolve into by $z=0$, according to simulations (\citealp{2013ApJ...779..127C}; \citealp{2015MNRAS.452.2528M}). 

The structure of this paper is as follows: in Section 2, we describe our observations, the photometric data, reduction and calibration; Section 3 includes source extraction and survey limits; Section 4 contains our method of selecting LAE candidates; Section 5 contains our main results on the region LAE overdensity, analysis and discussion; and our conclusions are presented in Section 6. Throughout this paper, we consider the flat $\Lambda$CDM universe model with a Hubble constant H$_{0}$ = 70 km s$^{-1}$ Mpc$^{-1}$, total matter density $\Omega_m$ = 0.3, and dark energy density $\Omega_{\Lambda}$ = 0. All magnitudes are presented in the AB system.

\begin{figure}
\centering
\includegraphics[scale=0.4]{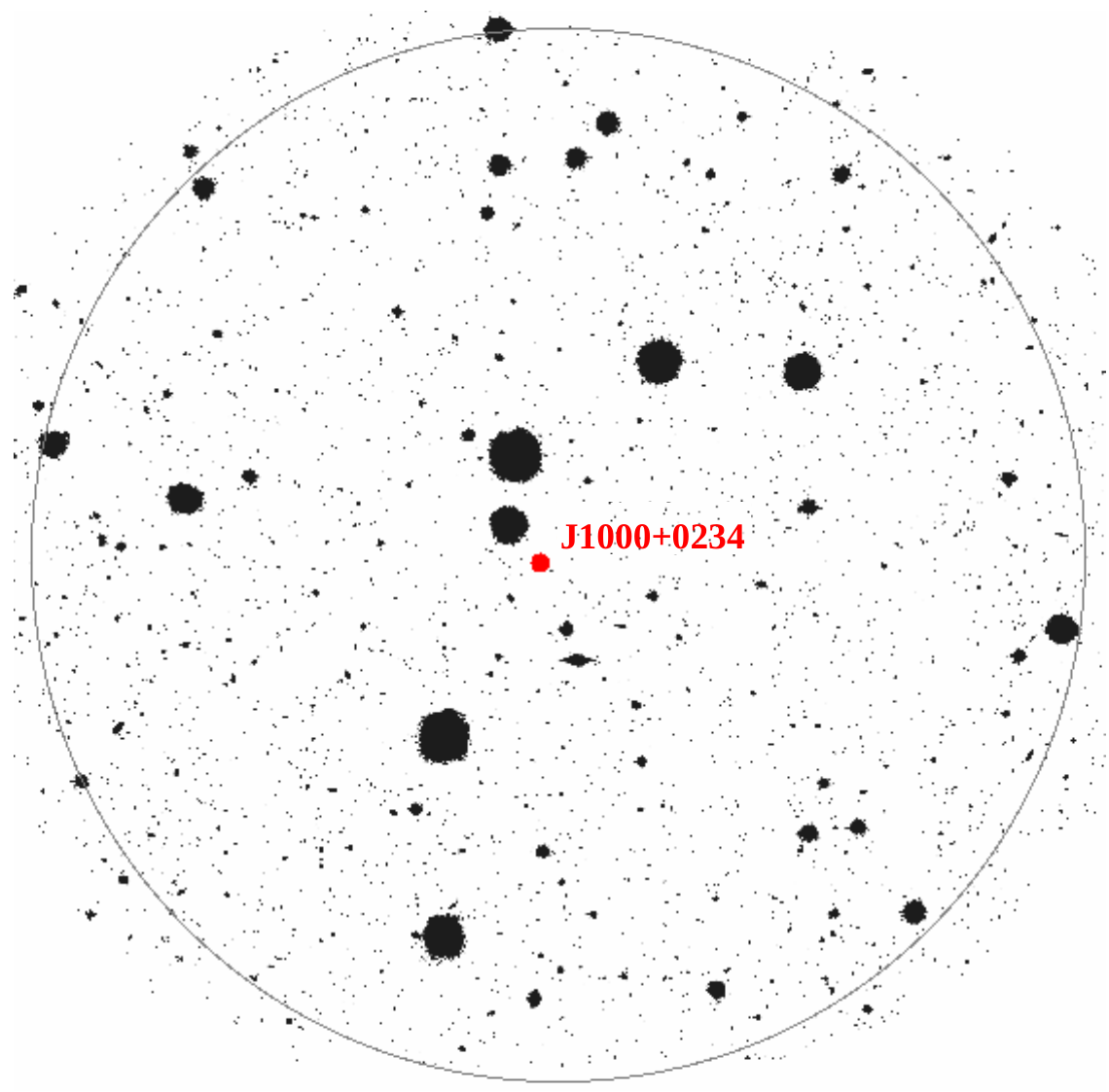}
\caption{IMACS image, obtained with the narrow-band filter ($\lambda_{c} = 6760 \text{Å}$), of a section of the COSMOS field, centered on the coordinates of the submm source J1000+0234 (also known as AzTEC/C17), (RA, DEC) = (10:00:54,+02:34:35), marked by the red circle at the center of the field. The image shows a field of view (FOV) of 27.4 arcmin for the IMACS camera.}
\label{fig:campo}
\end{figure}

\section{Photometric data, reduction and calibration}

We obtained 
photometric narrow-band data with the Inamori-Magellan Areal Camera \& Spectrograph (IMACS) on the Baade telescope at Las Campanas Observatory over four nights of observations. We observed a circular 0.16 deg² section of the COSMOS field centered at coordinates (RA, DEC) = (10:00:54,+02:34:35),  where the submm source J1000+0234 \citep{Schinnerer_2008,Scott_2008,Capak_2008}  
is located (see Figure \ref{fig:campo}). Additionally, we used broad-band photometric data from Suprime-Cam in the r+ filter over the same region, available on the COSMOS survey website.

\begin{figure}[b]
\centering
\includegraphics[scale=0.6]{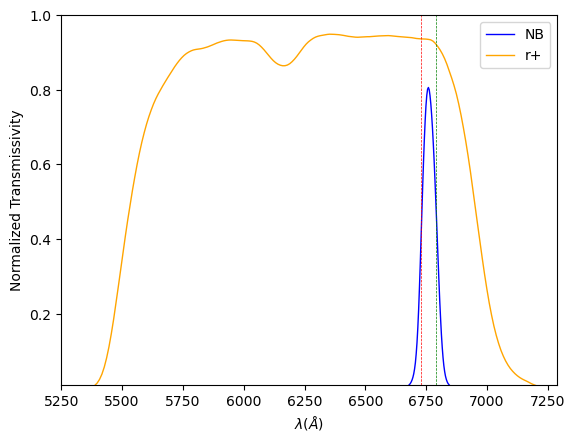}
\caption{Transmission curves of narrow-band filter (in blue) and r+ filter from SuprimeCam (in orange). The vertical dotted lines mark the limits of the full width at half maximum (FWHM) of the narrow-band filter, centered at 6760\AA, which correspond to the observed Ly$\alpha$ wavelength when emitted at the redshifts $z= 4.51$ and $z= 4.57$.}
\label{Fig:transmissivity}
\end{figure}

\subsection{Narrow-band data}

The IMACS detector is composed of eight CCDs that form a mosaic of 8192 $\times$ 8192 pixels. With a pixel scale of 0.2”, the FOV results in a footprint of 
$\sim$ 60 cMpc in 
diameter at $z\sim$ 4.5, centered on J1000+0234. The field was imaged for a total integration time of 13.6h using a narrow-band filter (centered at 676 nm, FWHM = 6.5 nm), which allows the detection of the Ly$\alpha$ emission line within a redshift range of $\Delta z\sim 0.06$ centered on $z= 4.54$ 
(see Figure \ref{Fig:transmissivity}). 




We carry out the narrow-band NB image reduction using a custom Python-based pipeline. Each IMACS exposure generates 8 \texttt{FITS}-formatted images, one for each of the eight detector chips. We reduce the chip images individually, according to the night of observation. Following the guidelines provided in the \href{https://www.lco.cl/technical-documentation/imacs-user-manual/}{IMACS user manual}, we subtract the bias using the overscan region. We apply flatfield correction by normalizing all science images with a masterflat 
to correct for pixel-to-pixel sensitivity variations.

Subsequently, we apply masks to eliminate cosmic rays in individual science images using \texttt{L.A.Cosmic} \citep{LA_COSMIC} and mask defective pixels using the \texttt{ccdmask} function from the \texttt{ccdproc} package \citep{ASTROPY}. We create individual chip masks for pixel regions shadowed by devices obstructing light within the camera, such as guides and deflectors. We then mask the edges of the science images and obtain astrometric solutions for each one using \texttt{Astrometry} \citep{ASTROMETRY} and \texttt{SCAMP} \citep{SCAMP}. All images are 
combined into a single NB mosaic using \texttt{SWarp} \citep{SWARP}.

\begin{figure}[b]
\centering
\vspace{2.2cm}

\begin{minipage}{0.46\textwidth}
    \centering
    \vspace{-1.6cm}
    \includegraphics[scale=0.6]{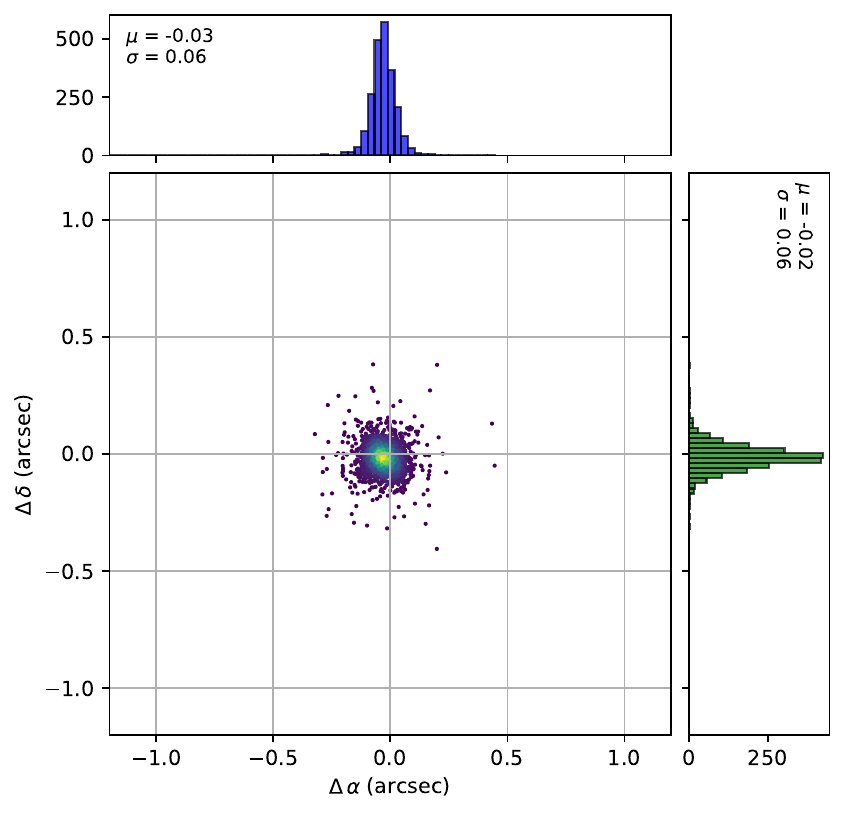}
    \caption{Astrometry verification: Difference between COSMOS2020 \citep{COSMOS2020} and NB sky positions, along with the histograms for $\Delta \alpha$ and $\Delta\delta$, for 2243 matches. We obtain a mean value of $\Delta\alpha = 0.02$'' and $\Delta \delta = 0.016 $'', with both standard deviations of 0.06''.}
    \label{FIG:ASTROMETRIA}
\end{minipage}%
\hfill
\vspace{1.7cm}
\begin{minipage}{0.48\textwidth}
    \centering
    \includegraphics[scale=0.55]{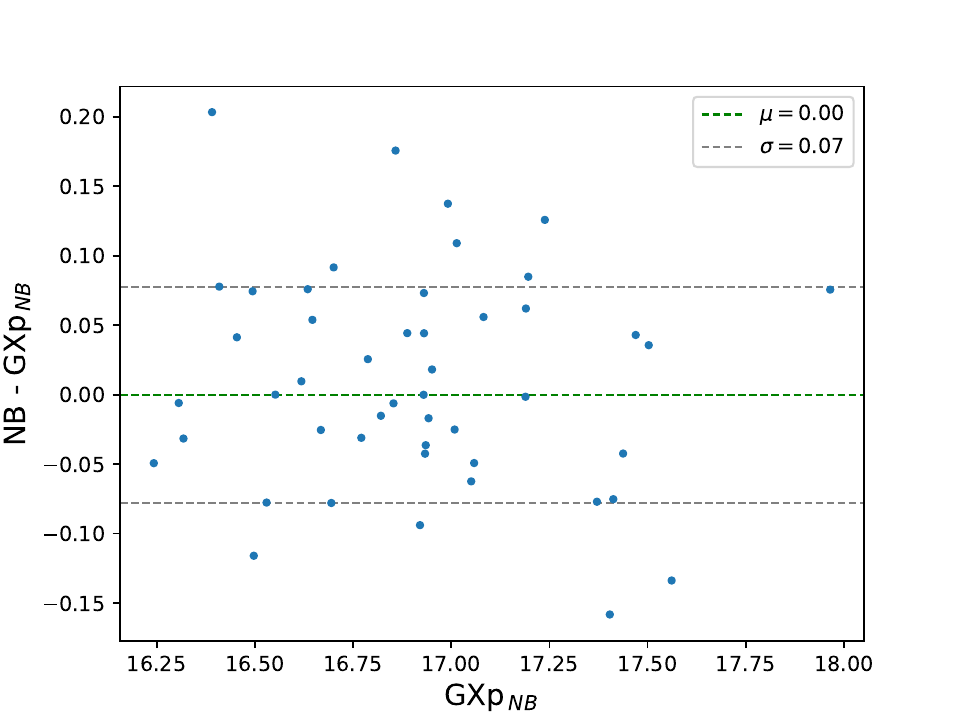}
    \caption{Photometric verification: Difference between our calibrated NB magnitudes (NB) and GaiaXPy synthetic NB magnitudes (GXp$_{NB}$). We note that 83\% of the sources have magnitude offsets smaller than 0.1 mag, and 50\% smaller than 0.05 mag. The zero-point (ZP) error, estimated as the standard deviation of the offsets, is 0.07 mag.}
    \label{FIG:ZP_MAG_DISP}
\end{minipage}

\end{figure}

 In order to verify the astrometry of the final NB image, we run SExtractor \citep{SExtractor}, selecting sources 
with \texttt{CLASS\_STAR} = 0.9 and a signal-to-noise ratio (S/N) between 5 and 1000,   
and compare the sky positions of the sources in the NB catalog with those in the COSMOS2020 catalog \citep{COSMOS2020}, using the r+ data, astrometrically aligned to Gaia DR1 \citep{Gaia_Lindegren_2016}. 
Figure \ref{FIG:ASTROMETRIA} shows the difference between the NB sky positions relative to COSMOS2020 sky positions, along with the histograms for $\Delta \alpha$ and $\Delta \delta$ for 2243 matches. We obtain mean values of $\Delta \alpha = 0.02$'' and $\Delta \delta = 0.03$'', where both distributions have a standard deviation of 0.06''.

\subsubsection{Narrow-band photometric calibration}

We use the python toolkit \texttt{GaiaXPy} \citep{2023_GaiaCollaboration} to perform the photometric calibration of our NB data. \texttt{GaiaXPy} uses Gaia low-resolution spectra to generate synthetic photometry based on the transmission curves of the filters of interest. 
We select sources with \texttt{CLASS\_STAR} = 0.9, \texttt{FLAG} = 0 and a signal-to-noise ratio (S/N) between 5 and 1000. 
We determine the zero-point magnitude using 48 stars for which the \texttt{GaiaXPy} mean spectra are available. We apply an aperture correction to our instrumental magnitudes obtained with a 1.3'' diameter aperture, which maximizes the S/N for our sources, to account for the flux lost outside the photometric aperture.
To calculate for the flux lost outside the photometric aperture we built a detailed Growth curve for the narrowband filter based on the instrumental magnitude in 30 different fixed, circular apertures ranging from 0.8" to 15.3" in diameter, with consecutive apertures differing by 0.5" in diameter, $a_{n+1} = a_n + 0.5"$. The correction consists of adding the cumulative mean instrumental magnitude difference between consecutive apertures, from the 1.3" aperture to the 14.3" aperture, where the increase in mean instrumental magnitudes between consecutive apertures becomes less than 0.001. See Figure \ref{FIG:APER_correction} for details.


We calculate the zero-point magnitude (ZP) as the median value of the difference between aperture corrected instrumental NB magnitudes (NB) and \texttt{GaiaXPy} magnitudes ($GXp_{\,NB}$):
\begin{equation}
ZP=median(NB - GXp_{\,NB})
\end{equation}
Figure \ref{FIG:ZP_MAG_DISP} shows the offsets between our calibrated NB magnitudes and \texttt{GaiaXPy} synthetic NB magnitudes, after the zero-point correction. A total of 83\% of the sources have magnitude differences smaller than 0.1 mag, and 50\% smaller than 0.05 mag. The zero-point (ZP) error, estimated as the standard deviation of the offsets distribution, is 0.07 mag.

\subsection{Broad-band data}

\begin{figure}
\includegraphics[scale=0.6]{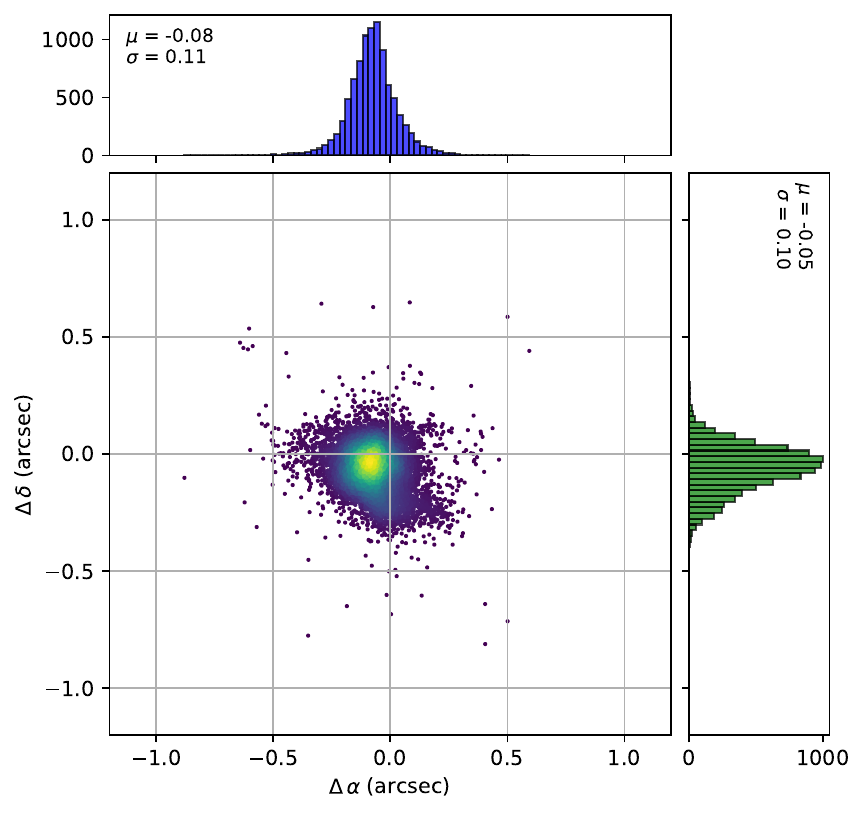}
    \caption{Astrometry verification: Difference between COSMOS2020 \citep{COSMOS2020} and r+ sky positions, along with the histograms for $\Delta \alpha$ and $\Delta\delta$, for 10430 matches. We obtain a mean value of $\Delta\alpha = 0.08$'' and $\Delta \delta = 0.05 $'', with standard deviations of 0.11 and 0.10'', respectively.}
    \label{FIG:COSMOS_offset}
\end{figure}%
\hfill
\begin{figure}[h]
    \includegraphics[scale=0.6]{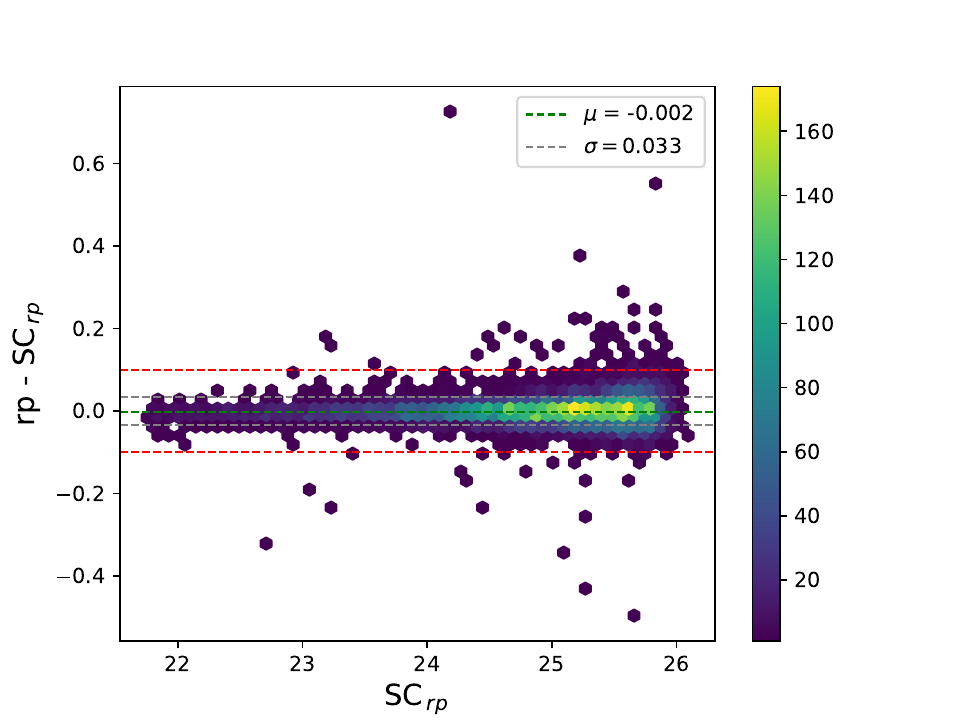}
    \caption{Photometry verification: Difference between our calibrated r+ magnitudes (rp) and the values in COSMOS2020 catalog \citep{COSMOS2020}  (SC$_{\,rp}$). 76\% of the sources have magnitude offsets smaller than 0.1 mag (red dashed line), and 71\% smaller than 0.05 mag (gray dashed line). The r+ mag error, estimated as the standard deviation of the magnitude offsets, is 0.033 mag.}
    \label{FIG:COSMOS_mag_offset}
\end{figure}

The Cosmic Evolution Survey (COSMOS, \citealp{COSMOS_survey}) is a vast extragalactic survey designed to explore the 
evolution of galaxies 
across the redshift range $z>$ 0.5 - 6. Spanning 2 deg$^2$ on the sky, privileged accessibility due
to its equatorial location, COSMOS incorporates multiwavelength imaging and spectroscopy from X-ray to radio wavelengths, provided by a wide range of instruments (\citealp[e.g.,][]{Aretxaga_2011_submm,Oliver_2012_COSMOS,Civano_2016_COSMOS,Marchesi_2016_COSMOS, Miettinen_2017_COSMOS, Aihara_2019_COSMOS, LeFevre_2020_COSMOS}), 
making it an exceptional resource for investigating the nature and evolution of galaxies across cosmic history.
In this work we use deep broad-band r+ images from the SuprimeCam to ensure the extraction and estimation of r+ magnitudes, even for NB sources with faint continuum. The r+ filter is centered at the wavelength 6305\AA, with a width of 1376\AA. 
We use SWarp \citep{SWARP} to combine the individual r+ images 
into a mosaic covering the entire section probed by our team. We use SCAMP \citep{SCAMP} to refine the astrometric solution of the r+ mosaic.
In order to use the NB image as a detection image to extract source magnitudes, 
we reprojecte the r+ image onto the narrow-band WCS grid. 
We refine the astrometric solution using SCAMP. To verify the astrometry and photometry, we select sources with SExtractor \texttt{CLASS\_STAR} = 0.9, FLAG = 0 and 5 $<$ S/N $<$ 1000. 
We compare their sky positions with those provided by the official COSMOS2020 catalog \citep{COSMOS2020}. Figure \ref{FIG:COSMOS_offset} displays the right ascension and declination offsets between the COSMOS2020 catalog and reprojected r+ image for the 10430 sources. We obtain mean values of $\Delta \alpha = 0.08$'' and $\Delta \delta = 0.05$'', with standard deviations of 0.11 and 0.10'', respectively. The mean offset in RA and DEC is on the order of the pixel size difference between the original and reprojected images, 0.05''. The reprojection may introduce small astrometric errors but still maintains a sufficiently good astrometric precision, smaller than the pixel size, 0.2''.

We rescale the r+ image flux to match the same narrow-band ZP magnitude and check the photometry accuracy by comparing the magnitudes of the flux rescaled r+ image with the 2'' aperture r+ magnitudes provided by the official COSMOS2020 catalog. Figure \ref{FIG:COSMOS_mag_offset} displays the photometric precision of the r+ flux rescaled image. A total of 76\% of the sources have magnitude offsets smaller than 0.1 mag, and 71\% smaller than 0.05 mag. The r+ mag error, estimated as the standard deviation of the distribution of magnitude offsets between our own r+ fluxes and those published by the COSMOS collaboration, is 0.033 mag.

\begin{figure}[b]
\centering
\includegraphics[scale=0.25]{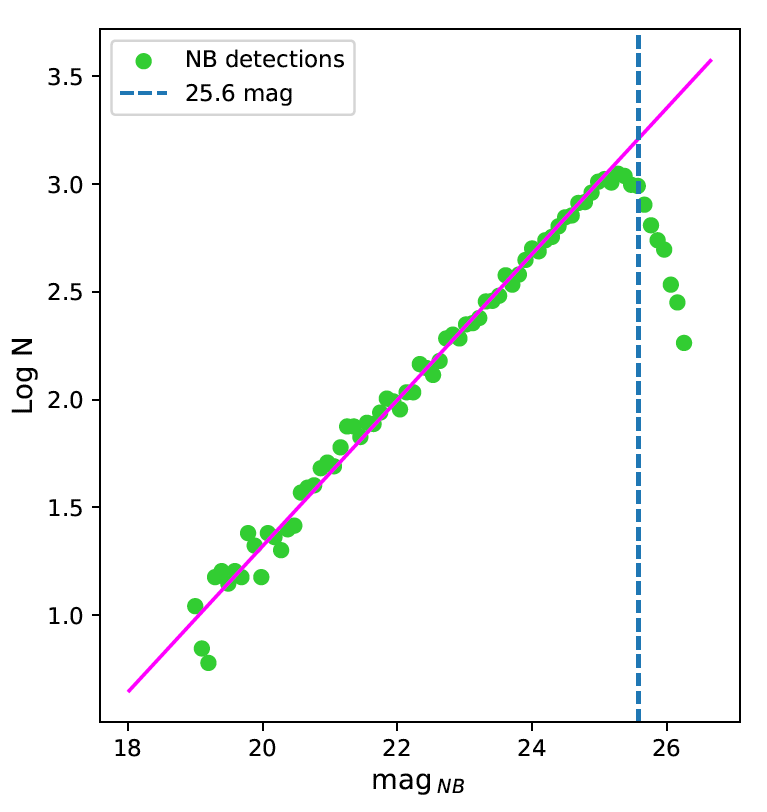}
\caption{
Logarithmic plot of the number of NB detections. The magenta line represents a linear fit given by y = 0.34 x - 5.45. 
At mag$\,_{NB}$ = 25.6, the sample shows a completeness of 60\%, estimated by taking the ratio of the number of NB sources detected in the sample (979) over the number of sources expected based on the linear fit (1204 sources)}.
\label{FIG:completness}
\end{figure}

\section{Source extraction and survey limits}
\label{sec:mag_lim}
Considering that LAEs are commonly young galaxies and may display a weak continuum, we use the NB image as a detection image in SExtractor \citep{SExtractor} dual-image mode to measure the color of the sources. 
We use a fixed aperture of 2'' diameter to measure the magnitudes of sources in the NB and r+ images. The detection threshold is set to select sources having at least four adjacent pixels above 1.5 $\times$ rms of the local background. This configuration yields a contamination rate by spurious sources of less than 0.3\%. The magnitude detection limit at a 5$\sigma$ depth for the narrow-band (NB) image is 25.6 mag. At this magnitude, our sample has a completeness of 60\% (see Figure \ref{FIG:completness}),  which we adopt as the cut-off in this work. We also adopt a cut at S/N $>$ 5, ensuring  high-quality object detection, and exclude objects brighter than mag$_{NB}$ = 19 to avoid saturated sources. We excluded sources detected solely in the NB image, as they may be spurious detections caused by noise spikes. This decision is supported by the assumption that, for Gaussian-distributed noise, fluctuations around the background level should occur symmetrically in both positive and negative values. Indeed, the number of sources detected in the inverted NB image - which consists purely of noisy features - closely matches the number of NB-only detections, evincing their genuinely uncertain nature. For the r+ image, the magnitude detection limit at a 3$\sigma$ depth is 27.1 mag. Our sample consists of 18413 sources within the 0.16 deg$^2$ area probed.


\section{\texorpdfstring{Ly$\alpha$}{Ly alpha} emitter candidates selection}

\subsection{Color excess}
\label{sec:color_excess}

\begin{figure*}
\centering
\includegraphics[scale=0.5]{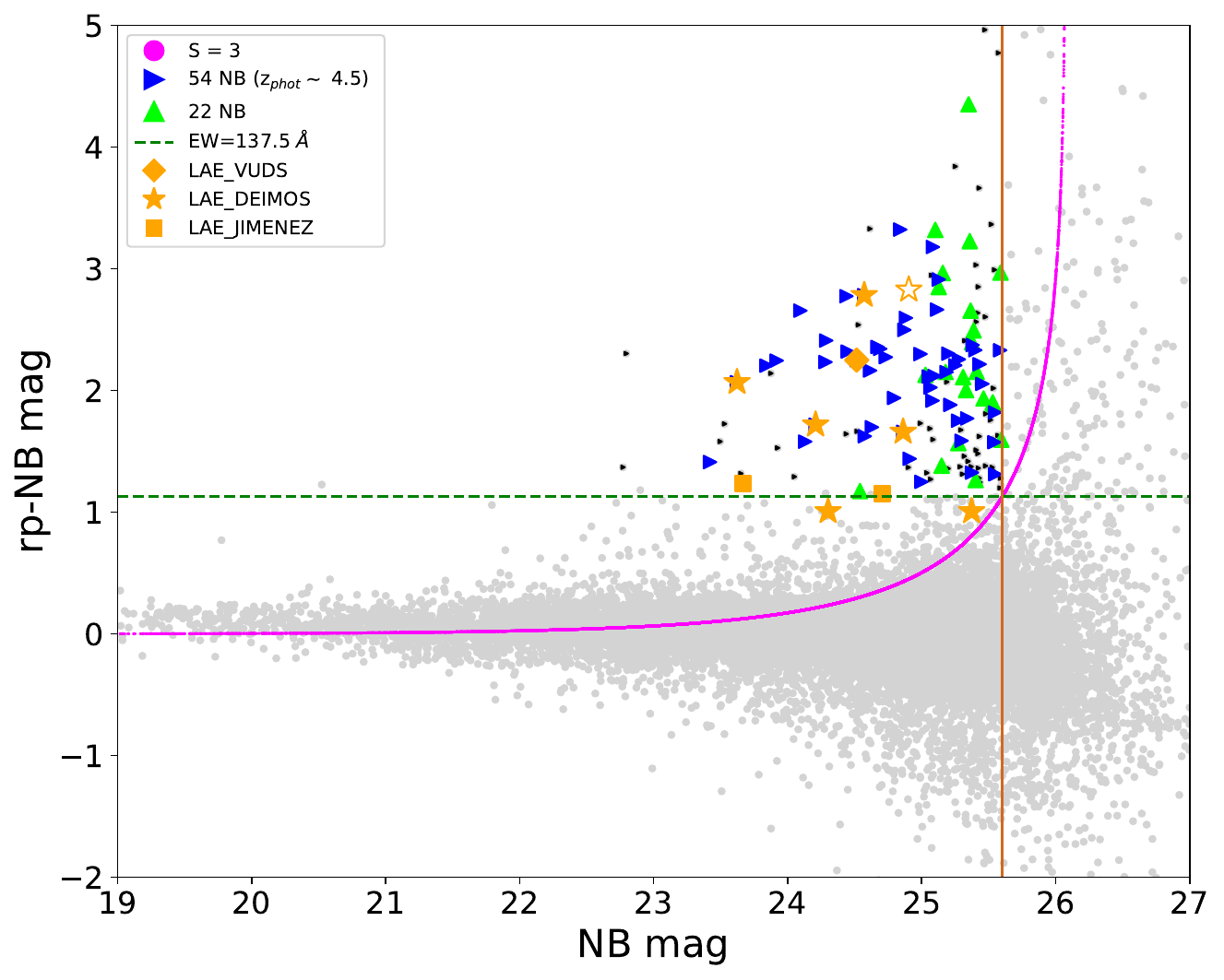}
\caption{Color-magnitude diagram of all detected sources in our field. LAE candidates are marked with colored symbols. Blue triangles are LAE candidates with photometric redshifts from COSMOS2020 catalog \citep{COSMOS2020} at $z_{phot}$ $\sim$ 4.5 and green triangles are LAE candidates with undetermined redshift. 
All 9 spectroscopically confirmed LAEs are indicated by filled orange symbols: stars are spectroscopic detections from DEIMOS 10k catalog \citep{DEIMOS_10k}, squares from \cite{Jimenez-Andrade_2023} and diamond from VUDS catalog \citep{Le_Fevre}. One spectroscopic detection from  DEIMOS 10k falls within a masked region and is represented by a hollow star. Black dots are candidates excluded by their photo-z or cutout analysis. The horizontal dashed line marks the cut in EW $>$ 137.5 \AA, while the vertical one represents the cut in NB magnitude at 25.6. The magenta curve represents the $3\,\sigma$ NB excess cut.
}
\label{FIG:ColorxMag}
\end{figure*}

In order to identify sources with candidate emission lines based on photometry, it is necessary to estimate the flux excess measured by the NB filter relative to the continuum, obtained from the r+ filter for each source. As the central wavelength of the NB is offset (see Figure \ref{Fig:transmissivity}) from the central wavelength of the overlapping r+ band and is located near the i+ band, we need to apply a correction to the continuum estimate of the sources. If only the r+ band were used to estimate the continuum, the total continuum flux would be the average over the r+ filter, leading to an underestimate of the true continuum flux underlying the Ly$\alpha$ emission, especially for sources with most of their emission located on the red side of the Ly$\alpha$ line. Similar corrections have been applied in previous NB surveys \citetext{e.g., \citealp{Sobral_2013,Vilella-Rojo_2015}; Sobral \citeyear{Sobral_2017,Sobral_2018}}. This correction ensures that the measured NB excess is not dependent on the intrinsic slope of the continuum and contributes to minimizing the detection of false emission lines. We perform this correction by analyzing the color dependence of \textit{rp - NB}, hereafter, on \textit{rp - ip} for sources with r+ and i+ counterparts in COSMOS2020 \citep{COSMOS2020} with good photometry (FLAG = 0 and S/N $>$ 5). We fit this relation with a linear model and use the coefficients, \textit{a} and \textit{b}, to correct the colors:
\begin{equation}
rp - NB = (rp - NB)_0 - a RI - b,
\end{equation}
where \textit{RI} is the color \textit{(rp - ip)} obtained from the COSMOS2020 catalog. For sources that do not have r+ or i+ detections in the COSMOS2020 catalog, we apply the median color correction of -0.01. With this correction, the color dispersion in the color-magnitude diagram is smoothly reduced. Figure \ref{FIG:ColorxMag} shows the resulting color-magnitude diagram used for our selection of LAE candidates. 

If a source has a strong emission line within the coverage of the narrow-band (NB) filter, it will have a higher flux density (erg/cm$^2$/s/Hz) in the narrow-band image compared to the broad-band (r+) image. This results in positive colors in the color-magnitude diagram. If a source does not exhibit an emission line within the NB coverage it will result in a \textit{rp - NB} color approximately zero. In order to select genuine emission line sources, specially Ly$\alpha$ emitters (LAEs), we apply two additional constraints for source selection, which are:\\

(i) Observed equivalent width (EW) $\geq$ 137.5\AA. This threshold corresponds to the rest-frame EW$_0$ $\gtrsim$ 25\AA, following other narrow-band surveys of LAEs \citep[e.g.][]{Ouchi_2008}, to minimize contamination from other line-emitters, as Ly$\alpha$ typically exhibits the highest observed EW. 
Recent surveys explored lowering this cut and showed that although a few additional real Ly$\alpha$ sources populating the bright end of the luminosity function may be recovered  \citep[e.g.,][]{Sobral_2017}, lower EW cuts also introduce a greater proportion of lower redshift contaminants. 
We implement this EW cut by 
considering the \textit{rp - NB} colors of our 
galaxies, as given by:
\begin{equation}
rp - NB =-\,2.5\,\log\left( \frac{1+EW/\Delta\lambda_{rp}}{1+EW/\Delta\lambda_{NB}} \right)
\end{equation}

where \textit{NB} and \textit{rp} are the apparent magnitudes in the narrow-band (676nm) and broad-band (r+) filters, $\Delta\lambda_{NB}$ and $\Delta\lambda_{rp}$  the widths of the filters —  65\AA \, and 1376 \AA\, for the narrow-band and the r+ filters, 
respectively (see \citealp{2004_Palunas}). An emission line with EW=137.5\AA \, produces a color of $\sim$ 1.13; we use this cut to clean our sample of narrow-band excess objects and better select Ly$\alpha$ 
emitters (see Figure \ref{FIG:ColorxMag}).\\

(ii)\; $ rp - NB >$ $S\,\sigma$, with $S$ = 3. Considering the sky as the sole source of noise, positive colors can arise due to scatter in the fluxes measurements. We assume the color $rp - NB =0$ as a reasonable approximation for the measurement of the continuum of a high-redshift galaxy. The parameter $S$ is used to quantify the significance of the narrow-band flux excess relative to the random scatter expected for a zero-color source. 
The condition for a source to be considered a genuine narrow-band excess object is given by \citep{Bunker_1995}: 
\begin{equation}
    c_{NB}-c_{rp}=S \,\sigma\,,
\label{eq:counts}    
\end{equation}
where c$_{NB}$ and $c_{rp}$ correspond to the counts in the NB and rp images, respectively, and $\sigma=\sqrt{\pi r^2(\sigma_{NB}^2+\sigma_{rp}^2)}$ is the photometric error determined by the combination of the standard deviations of counts within the photometric aperture (with radius r) used for flux measurements in the broad-band rp and NB images. For images with the same zero-point magnitude, equation (\ref{eq:counts}) can be expressed in terms of magnitudes as
\begin{equation}
   rp - NB= -2.5 \log_{10} \left( 1 - S \,\sigma \, 10^{-0.4(ZP - NB)} \right).
\label{eq:fot_scatter}
\end{equation}
\vspace{0.01cm}

Considering these criteria, the sample of genuine narrow-band excess 
objects can be identified in Figure \ref{FIG:ColorxMag} as the colored data points above the magenta curve, which corresponds to the condition defined by equation \ref{eq:fot_scatter}. Our color criteria result in a total of 147 LAE candidates.

These sources are identified 
as “candidates” considering 
that interlopers at low redshift can also be detected as narrow-band excess objects under the above conditions, although the cut at EW=137.5 already excludes a large number of the bright low-redshift sources. These consist primarily of [OII] emitters at $z\sim0.8$, [OIII] emitters at $z\sim0.3$, and H$\alpha$ emitters at $z\sim0.03$.

\subsection{Photometric redshifts of line emitters}
\label{phot-z line emitters}
\cite{Smol_2017}, in a search for high-redshift overdensities in the COSMOS2015 catalog, evaluated the quality of the photometric redshifts from \cite{Laigle_2016} by comparing photometric redshifts (3.5 $\leq$ $z_{phot}$ $\leq$ 6) with 240 secure spectroscopic redshifts in the COSMOS field. They fit a Gaussian to the ($z_{phot}$ - $z_{spec}$) / (1 + $z_{spec}$) distribution and verified the catalog's accuracy, finding a strong agreement with a standard deviation of $\sigma_{\Delta z/ (1 + z)}$ = 0.0155 for sources with mag$_{\,i+}$ $<$ 25.5 (Figure 1 of \citealp{Smol_2017}). However, due to misidentification between the Lyman and Balmer breaks in the observed SEDs, some high-redshift sources were misidentified as low-redshift \citep[see][]{Le_Fevre_2015}. They estimated a sample completeness of 80\% for sources with $z_{spec}$ $>$ 3.5.

The COSMOS2020 catalog \citep{COSMOS2020}, used in this work, has achieved an improvement in photometric redshift precision of approximately an order of magnitude compared to the previous COSMOS2015 catalog \citep[see section 7,][]{COSMOS2020}. This enhancement makes the catalog a reliable source for photometric redshifts and suitable for the identification of high-redshift overdensities. 
Spectroscopic observations of our LAE candidates are still required to confirm the nature of the sources, especially of the fainter ones, which are associated with greater uncertainties in photometric redshift estimates.

\begin{figure}
\centering
\includegraphics[scale=0.28]{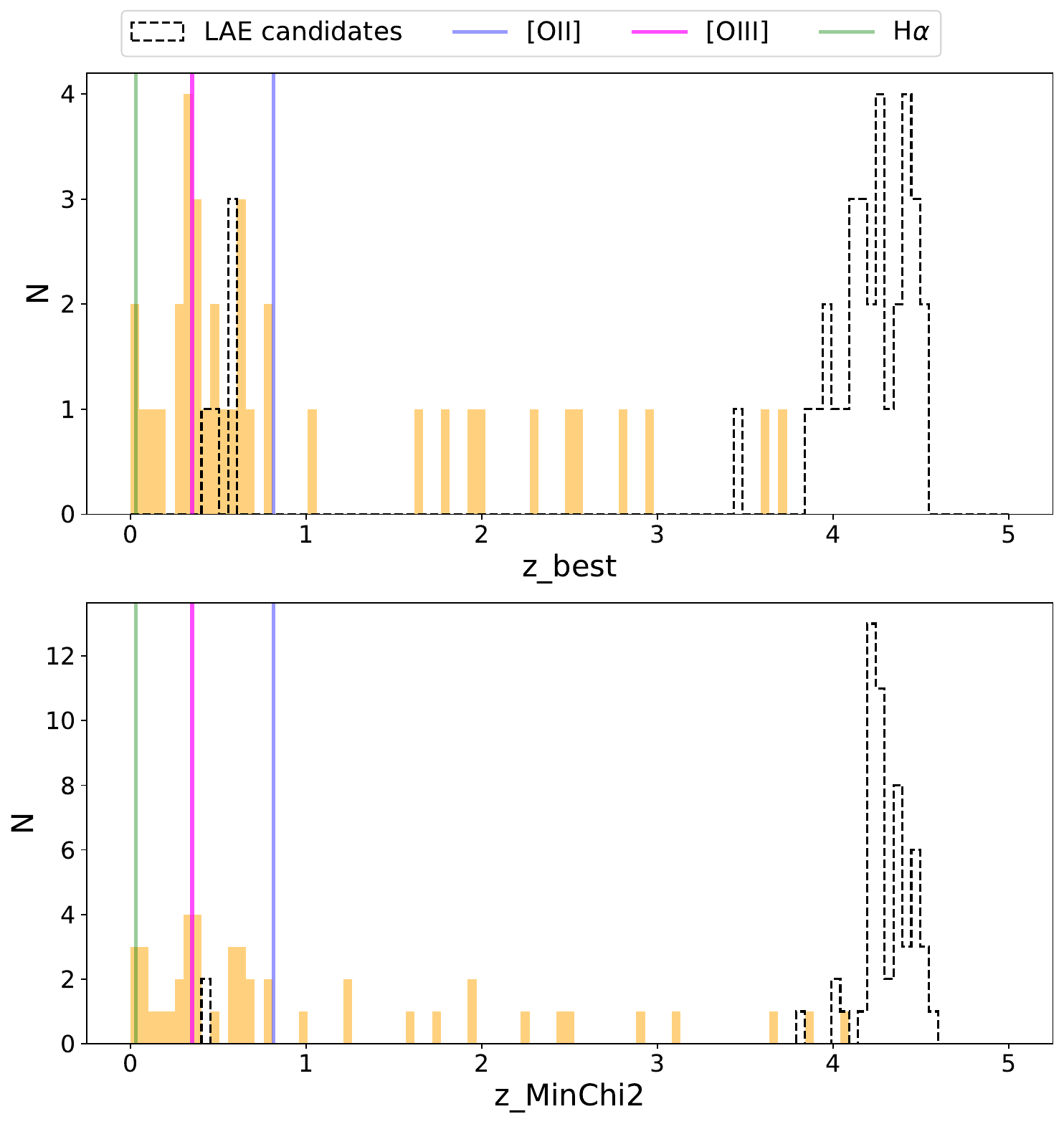}
\caption{Photometric redshift distribution of 99 line emitters with matches in the COSMOS2020 catalog by \cite{COSMOS2020}. Colored lines show the expected redshift for sources emitting O[II], O[III] and H$\alpha$ lines.  (Upper): Redshift distribution relying on z\_BEST. The 5 LAE candidates with 0 $<$ z\_BEST $<$ 1 also have the alternative photo-z solution z\_MinChi2 $\sim$ 4.5. Among low-redshift interlopers there is a prominent peak consistent with [OIII] line emission at z\_BEST $\sim$ 0.3. (Bottom):
Redshift distribution relying on z\_MinChi2. We see peaks of at least three sources consistent with [OIII] and H$\alpha$ line emissions.  
In total, 54\% of the sources 
are classified as LAE candidates and 46\% 
as low-redshift interlopers.  
}
\label{FIG:excluidas}
\end{figure}

We crossmatched our LAE candidate catalog with the COSMOS2020 catalog in the r band to select the sources at $z\sim$ 4.54 and eliminate low-z interlopers. The catalog offers the deepest data in the COSMOS field for $\sim$ 1.7 million objects. For consistency with other works analyzing the same sky region and its surroundings \citep{Smol_2017, Crespo_2021, Jimenez-Andrade_2023, Staab_24_Taralay}, we adopt the CLASSIC version of the catalog and the photometric redshifts determined by the Le Phare photo-z code \citep{Arnouts_2002_phot-z,Ilbert_2006_phtoz}. The precision of photometric redshifts is approximately $\sigma_{\Delta z/ (1 + z)}$ = 0.01 for sources with magnitudes brighter than mag$\,_i$ $<$ 22.5. A small decrease in precision is observed at fainter magnitudes $ \sigma_{\Delta z/ (1 + z)}$ =  0.025 for sources with mag$\,_i$ $<$ 25. For sources with  mag$\,_i$ $<$ 27, the precision is on the order of $\sigma_{\Delta z/ (1 + z)}$ = 0.044 \citep{COSMOS2020}. We used $\sigma_{\Delta z/ (1 + z)}(1 + 4.5)$ for the photo-z uncertainty 
and searched for LAE candidates within the wavelength range covered by the NB filter,  corresponding to the redshift range 
between $z_{min} - 1.5 \sigma_{\Delta z/ (1 + z)}(1 + 4.5)\,<\, z_{phot} \,<\, z_{max} + 1.5 \sigma_{\Delta z/ (1 + z)}(1 + 4.5)$, with $z_{min} = 4.51$ and $z_{max} = 4.57$. To select our sources we used z\_BEST, the photo-z derived using a method similar to \cite{Ilbert_2009, Ilbert_2013}, and z\_MinChi2, the photo-z measured using galaxy templates as the minimum of the $\chi^2$ distribution. From the 147 LAE 
candidates, we find 1” matches for 99 sources; Figure \ref{FIG:excluidas} shows the COSMOS2020 photo-z distribution for these 99 sources. Of these, 
46 sources have lower-redshift COSMOS2020 photo-zs, representing 46\% of the matches. With this, we reduce our sample of LAE candidates around J1000+0234 to 101 sources.

\subsection{Cutouts inspection}
\label{subsec:cutouts}
We visually inspect image cutouts for the 101 LAE candidates to eliminate false detections, such as escaping light from saturated pixels, artifacts or sources in noisy regions. A total of 25 sources were excluded from the candidate sample. This reduced our sample to 76 genuine candidates. Among these, 54 candidates exhibit photometric redshifts at $z\sim$ 4.54 and 22 candidates have unknown photometric redshifts, as they did not have a counterpart in COSMOS2020 \citep{COSMOS2020}. 
The cutouts of the 22 sources without matches in COSMOS2020 are shown in Appendix \ref{appendix:cutouts}.  %

\subsection{Spectroscopic confirmations}

To confirm our LAE candidates selection method, we use two spectroscopic catalogs from the literature: VIMOS Ultra-Deep Survey \citep{Le_Fevre_2015} and DEIMOS 10k Spectroscopic Survey \citep{DEIMOS_10k}, and spectroscopic detections made 
by \cite{Jimenez-Andrade_2023}. \cite{Lemaux_2018} used the VIMOS Ultra-Deep Survey catalog to discover the protocluster PCl J1001+0220 at $z\sim$ 4.7. \cite{Staab_24_Taralay} targeted this structure and reestablished its extent and internal structure at 4.48 $<$ $z<$ 4.64, renaming it Taralay protocluster. For this mapping, they used the data from the DEIMOS 10K Spectroscopic Survey catalog, and from the \href{http://www.orelsesurvey.com/c3vo.html}{Charting Cluster Construction with VUDS and ORELSE (C3VO)} survey, 
an ongoing imaging and spectroscopic campaign aiming to map the growth of structures up to $z\sim$ 5. 
\cite{Jimenez-Andrade_2023}, using the Multi Unit Spectroscopic Explorer (MUSE) to target the Ly$\alpha$ blob around the source J1000+0234, serendipitously detected 3 LAEs in a redshift bin $\Delta z\leq 0.007$ (i.e. $\lesssim$ 380 km s⁠$^{-1}$) located at $\lesssim$ 140 Kpc from the source.

We spectroscopically confirm 7 of our color-selected LAE candidates with photometric redshifts at $z\sim$ 4.5, within the range 4.52 $< z\,_{spec} <$ 4.58, using the spectroscopic data described above: We find 4 sources in the DEIMOS 10k catalog \citep{VIMOS}, 1 in the VIMOS catalog \citep{Le_Fevre}, and 2 in \citet{Jimenez-Andrade_2023} - corresponding to the Ly$\alpha$ blob ($L_{\mathrm{Ly}\alpha} = 41.17 \pm 0.31 \times 10^{43}$ erg s$^{-1}$) and the brightest LAE ($L_{\mathrm{Ly}\alpha} = 3.65 \pm 0.1\times 10^{42}$ erg s$^{-1}$), whose luminosity lies at the 95\% completeness limit of our sample. The other 2 LAEs reported in \citet{Jimenez-Andrade_2023} were too faint to be successfully detected by \texttt{SExtractor}.  We include these spectroscopically-confirmed sources in Figure \ref{FIG:ColorxMag}.

We also include in Figure \ref{FIG:ColorxMag} two additional spectroscopic detections from DEIMOS, although they fall below the  EW cut for our color selection. This demonstrates that lower EW limits can be applied to the color selection in order to include more sources, but we note that this is at the cost of significantly increasing the number of low-redshift interlopers. For this reason, we maintain our EW cut criterium to reduce errors in the LAE density estimate in the region.

\section{RESULTS AND DISCUSSION}
The Ly$\alpha$ emission is a tracer of galaxy formation, directly
connected to star formation events even in very young systems. Due to the low bias of LAEs \citep[see, e.g.][]{Gawiser_2007,Guaita_2010}, they can be used as faithful tracers of large-scale structures. In this work, we rely on the
identification of LAE candidates to
characterize an overdense region at
$z\sim$ 4.5 within the COSMOS field. Based on LAE numerical overdensity ($\delta$) relative to the field, we identify regions that could potentially evolve into galaxy clusters by $z=0$ and discuss their possible evolutionary pathways according to simulations. 
\subsection{Measurement of Volume Overdensity}
 The galaxy numerical overdensity ($\delta$) is defined as the volume density contrast compared to the field:
\begin{equation}
    \delta=\frac{n-n'}{n'},
\end{equation}
where $n$ is the volumetric number density of LAEs in the probed region
and $n'$ is the volumetric number density of LAEs in the field. A value of $\delta=0$ characterizes the field, $\delta>0$ an overdense region and $\delta<0$ an underdense region.

\subsubsection{Field LAE density }
\label{secao:fieldLAEdensity}
To calculate the LAE density in the 
field we first need to estimate the field luminosity function parameters at the redshift of interest, $z= 4.5$. This luminosity function, integrated down to the Ly$\alpha$ luminosity to 
which our survey is complete, provides the characteristic number density of LAEs 
expected down to the specific depth explored in our work. \cite{2020ARA&A..58..617O} provides a review on LAEs,
depicting the luminosity function of LAEs at redshifts 0.3, 2.1, 3.3, 3.7, 5.7, 6.6, 7.0, and 7.3 from a range of studies. We use interpolation to calculate the optimal parameters $L^\star$, $\Phi^{\star}$ and $\alpha$ for the Schechter luminosity function (Schechter 1976) 
 at $z\sim$ 4.5. We obtain $L^{\star}$ = 9.13 . 10$^{42}$ erg/s, $\Phi^{\star}$ = 3.76 . 10$^{-4}$ Mpc$^{-3}$ and
$\alpha$ = -1.8.

In order to calculate the Ly$\alpha$ line luminosity emitted by our LAE candidates, we first calculate the line flux following the method described in \cite{Sobral_2017}:
\begin{equation}
F_{\text{line}} = \frac{\Delta \lambda_{\text{NB}} \left( f_{\text{NB}} - f_{\text{rp}} \right)}{1 - \left( \Delta \lambda_{\text{NB}} / \Delta \lambda_{\text{rp}} \right)},
\end{equation}
where $f_{NB}$ and $f_{rp}$ are the flux densities measured within the narrow-band (NB) and broad-band (r+) filters, respectively. The line luminosity can be estimated as $L = 4 \pi D^2 F_{\text{line}}$, where $D$ is the luminosity distance.

\begin{figure}
\includegraphics[scale=0.42]{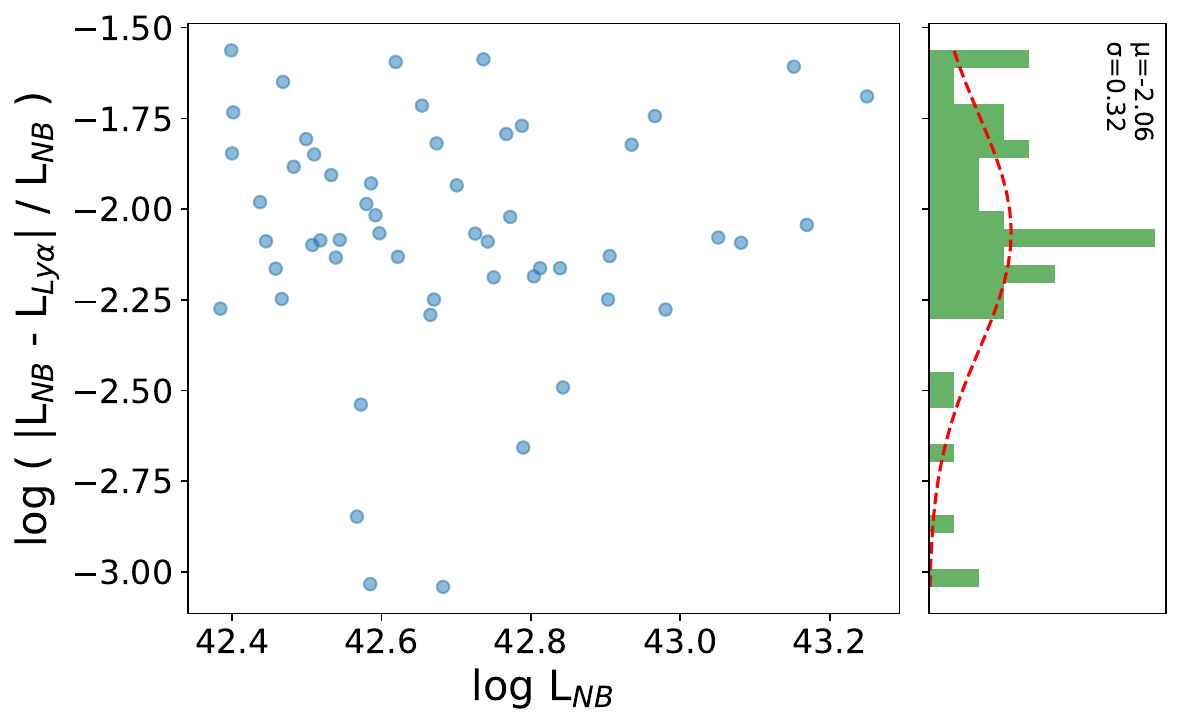}
\caption{Normalized difference between the NB and Ly$\alpha$ luminosities. On average, the difference between these luminosities is on the order of 0.01 of the NB luminosity. Considering the standard deviation of the Gaussian fit to the distribution, the variations are typically between 0.5\% and 2\% of the NB luminosity.}
\label{FIG:lumin_dif}
\end{figure}

Figure \ref{FIG:lumin_dif}
shows the normalized difference between the NB and Ly$\alpha$ luminosities. On average, the difference between these luminosities is on the order of 0.01 of the NB luminosity. Since the luminosities are closely matched, we assume that the completeness in the NB luminosity serves as a reasonable approximation for the completeness in the Ly$\alpha$ line luminosity.
 Our sample shows 60\% completeness at L$_{NB}$=$10^{42.37}$ erg/s (mag$_{NB}$ = 25.6) and 95\% completeness at L$_{NB}$=$10^{42.57}$ erg/s (mag$_{NB}$ = 25.1). Taking into account the percentage of sources missed within this magnitude range, we calculate the LAE field density we expect to observe by convolving the field LAE luminosity function with a function $f(L)$ (Figure \ref{Fig:function_f}) that accounts for the NB completeness as a function of luminosity: 
 \vspace{-0.5em}
\begin{equation}
    \bar{n}=\int_{L_i}^{L_f} \Phi(L) f(L)dL,
\end{equation}

where we integrate down to luminosity L$_{i}$=10$^{42.37}$ erg s$^{-1}$. We compute the LAE field density to be $(5.5 \pm 3.6) \times 10^{-4}\ \mathrm{cMpc}^{-3}$, based on integrating the luminosity function using a Monte Carlo approach \cite{Monte_Carlo_2020} that considers the uncertainties in $\phi^*$ and $L^*$. We also estimate the expected LAE field density using the luminosity function and its uncertainties derived from the SC4K sample of \cite{Sobral_2018}, obtaining $(3.35 \pm 3.04) \times 10^{-4}\ \mathrm{cMpc}^{-3}$. The two estimates are consistent, within the uncertainties, with the observational value of $(3.90 \pm 0.05) \times 10^{-4}\ \mathrm{cMpc}^{-3}$ reported by \cite{Herrera_2025}. We note that integrating luminosity functions inherently results in larger uncertainties compared to the method used by \citet{Herrera_2025}, which consists of a number count divided by the survey volume probed by their narrow-band N673 filter. We adopt the estimate based on \citet{2020ARA&A..58..617O} to ensure a more conservative approach, effectively setting an upper limit for the LAE field density.

\subsubsection{KDE density map}

The gaussian kernel density estimation is a non-parametric method frequently used in the literature to estimate a surface density of objects within a region of the sky (e.g., \citealp{Kyoung-Soo_2014}; 
\citealp{Badescu_2017};
\citealp{Hu_2021};
\citealp{2022_Huang}; \citealp{Ramakrishnan2023,Vandana_2024}). This method produces a smooth surface density map from the
discrete positions of the detected galaxies by smoothing over their positions using a Gaussian kernel.

\begin{figure*}
\hspace{1.5em}
\centering
\includegraphics[scale=0.48]{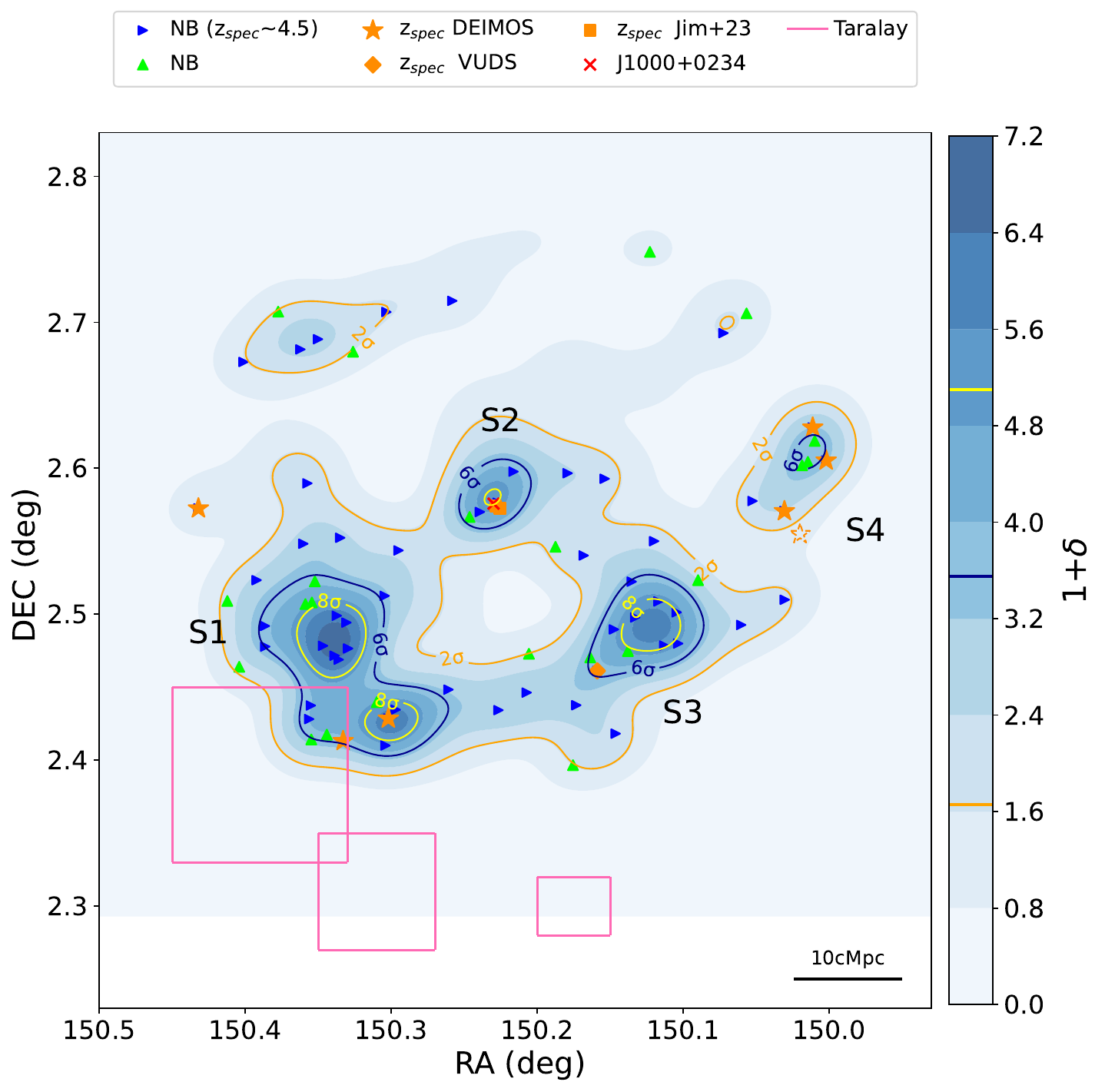}
\caption{Density map of our field displaying, the proximity between our galaxy overdensity, and the neighboring Taralay structure, shown as pink squares. Orange, indigo and yellow contours indicate LAE
overdensities of 2, 6 and 8$\sigma$ significance respectively. Values of (1 + $\delta$) are marked on the color bar. The four largest overdense regions are labeled by the names S1, S2, S3 and S4. LAE candidates are displayed in blue ($z_{phot}\sim$ 4.5) and green (undetermined $z_{phot}$) 
triangles. 
Spectroscopic detections are displayed in orange symbols: stars from DEIMOS 10K catalog \citep{DEIMOS_10k}, squares from \cite{Jimenez-Andrade_2023} and diamond from VUDS catalog \citep{Le_Fevre_2015}. J1000+0234 \citep{Schinnerer_2008,Scott_2008,Capak_2008} represented by a red cross at the center of the field. The three pink boxes show the location of Taralay protocluster \citep{Lemaux_2018, Staab_24_Taralay}.
}
\label{FIG:density_map}
\end{figure*}

Figure \ref{FIG:density_map} shows the density map of LAE candidates. We use the \texttt{gaussian\_kde} function from SciPy \citep{scipy} with a two-dimensional 
fixed-size kernel with FWHM of 0.19 deg, which corresponds to 25 cMpc in size at redshift $z\sim$ 4.5. This extension matches those of protoclusters at this redshift \citep[see, e.g.,][]{2015MNRAS.452.2528M}. Each time the Gaussian KDE code is run, data points are  randomly distributed  in order to fill the gaps left by removed sources inside the COSMOS masks for photometrically affected objects, ensuring masked regions match the field LAE density calculated in Section \ref{secao:fieldLAEdensity}. Additionally, to correct for potential low-redshift interlopers among LAE candidates with undetermined photo-zs, we use the fraction of low-redshift interlopers identified among sources with matches in the COSMOS2020 catalog (46\%), to randomly remove 46\%of the LAE candidates without photo-zs. We convert the surface density map into a volume density map by converting 1 $~\mathrm{deg}^2$ to the corresponding volume of $5.7 \times 10^5 ~\mathrm{cMpc}^3$ probed by the NB at  $z\sim$ 4.5. The overdensity map, depicting $1+\delta$ is computed by dividing the volume LAE density map by the volume LAE density in the field. We run the code 300 times, ensuring that the average density values remain consistent within 0.001 between successive runs, and adopt the mean value at each point on the map. 
 The colored curves represent the number of standard deviations in the density distribution: 2$\sigma$, 6$\sigma$ and 8$\sigma$. We calculate the value of $\sigma$ using an iterative sigma-clipping approach to prevent dense regions from biasing the value of $\sigma$. We fit a Gaussian to the distribution of the density values from the map and then iteratively clip outliers beyond 1.5 times the standard deviation. After each iteration, we recalculate the mean and standard deviation, continuing this process until the sigma value stabilizes at 0.0001, which was reached after 6 iterations.

Following \cite{Lemaux_2022}, we consider a 
protocluster region one that is delimited by an overdensity of log$(1 + \delta) = 0.22$,   
showing a peak of at least log$(1 + \delta)$ = 0.55.  
 The probed section of the COSMOS field displays 4 such regions (S1, S2, S3 and S4), marked by the $6\sigma$ (indigo) contours where log$(1 + \delta)$ = 0.55, satisfying these criteria. S1 is the largest and most extreme structure, spanning a region of 27 $\times$ 20 cMpc$^2$ out to its 2$\sigma$ outskirts, delimited by the orange contour where log$(1 + \delta) = 0.22$. It contains a dense core with $\delta > 4$, delimited by the 8$\sigma$ (yellow) contour. 
 This core has a radius of approximately 2.5 cMpc and harbors the highest overdensity on the map. 
S3 is the second densest region, spanning approximately 25 $\times$ 15 cMpc$^2$, considering its extension out to its 2$\sigma$ outskirts, also hosting a dense core with $\delta$ $>$ 4 delimited by the 8$\sigma$ line, with a radius of $\sim$ 1.7 cMpc. 

We find two other smaller overdense islands, S2 and S4. The region S2 hosts the submm source J1000+0234. This region reaches an overdensity peak of $\delta \sim$ 4 , with a radius of $\sim$ 7 cMpc out to its $2\sigma$ outskirts. The S4 region hosts 3 LAE spectroscopic detections from \cite{DEIMOS_10k}, showing a smaller peak overdensity of $\delta \sim$ 3, within an area of approximately 15 $\times$ 10 cMpc$^2$.





\subsection{Mass of the structure from Simulations}
\label{sec:mass from simulations}

\cite{2015MNRAS.452.2528M}, using a semi-analytic model based on the Millennium Simulation, highlights that the complex structure of protoclusters can lead them to be highly spatially extended. Consequently, even at large distances from the protocluster core, there may be dense regions populated by protocluster galaxies. Their study points that a protocluster evolving into the most massive cluster halos (M $\geq 10^{15}$ M$\odot$) spans a region with a radius of approximately 15 to 25 $h^{-1}$ cMpc at $z\sim$ 4. This region covers around 90\% of the structure's mass by $z=0$.
In Figure \ref{FIG:density_map}, the large structure composed by S1, S2 and S3, enclosed by the 2$\sigma$ contour, spans approximately 55 $\times$ 35 cMpc$^2$, which is consistent with the dimensions of massive clusters progenitors described by \cite{2015MNRAS.452.2528M}. This scenario is consistent with hierarchical structure formation models, where smaller structures merge over time to form more massive ones, such as clusters, by $z=0$. See \cite{Nicandro} for an example of a protocluster hosting two substructures, with a likelihood greater than 80\% of merging into a single cluster by $z\sim$ 0.1.
 
Taking a different approach, \cite{2013ApJ...779..127C} employs the dark matter Millennium Simulations to track the redshift evolution of structures that reach masses of $M \geq 10^{14}\,M_\odot$ by $z=0$ within volumes of $(15\,\mathrm{cMpc})^3$ and $(25\,\mathrm{cMpc})^3$.
They use different types of galaxies, including those with star formation rates greater than 1 M$_\odot$ yr$^{-1}$, that match the characteristics of LAEs. Their findings indicate that  a region with a volume of 15$^3$cMpc$^3$\,(25$^3$cMpc$^3$) and a numerical overdensity of $\delta$ = 2.83 (0.99) at $z= 4$ has an 80\% probability of forming a cluster by $z=0$ (see Table 4 in \citealp{2013ApJ...779..127C}). The numerical overdensity value provides an estimate of the mass that the structure will reach at $z=0$. However, \cite{2013ApJ...779..127C} argue that if the effective radius is significantly smaller than that expected according to the final mass of the structure (see Figure 2 in \citealp{2013ApJ...779..127C}), it will probably evolve into a lower-mass system.

S1 exhibits an average overdensity of $\bar{\delta}$ = 3 within a volume of (25 cMpc)$^3$, which is far greater than the threshold required to form a cluster with an 80\% probability.
If this entire region collapses to form a cluster by $z=0$, the effective radius calculated according to \cite{2013ApJ...779..127C} - see Appendix \ref{appendix:effective radius} for details - would be approximately 7 cMpc, consistent with median-mass clusters. Therefore, this overdensity has the potential to evolve into a cluster with a mass of 3 - 10 × 10$^{14}$ M$\odot$.
Similarly, S3 has an effective radius of approximately 7.5 cMpc and an average overdensity $\bar{\delta}$ = 3.8 within a volume of (15.7 cMpc)$^3$. This is consistent with the progenitors of low to medium-mass clusters at $z= 4.5$, with an estimated probability of $\sim$ 80\% (see Table 4 in \citealp{2013ApJ...779..127C}).


The region hosting the submm source J1000+0234, S2, shows an average overdensity of $\bar{\delta} = 2.9$ within a $(15 \,\mathrm{cMpc})^3$ volume, suggesting a 50–80\% probability of evolving into a low-mass cluster (M $< 3 \times 10^{14},M_\odot$) \citep[see Table 4 in][]{2013ApJ...779..127C}. However, this region requires a greater surrounding overdensity to continue growing and eventually form a cluster by  $z= 0$. The mapped overdense region, measured out to its 2$\sigma$ outskirts, has an effective radius of only $\sim1$ cMpc — just one-fourth of the expected size for low-mass cluster progenitors. As noted by \citet{2013ApJ...779..127C}, in the absence of such a large-scale overdensity, this region is more likely to be the progenitor of a much less massive structure, such as a group.  A similar case is seen in the less dense region S4, which shows $\bar{\delta} = 2.2$ within $(15 \,\mathrm{cMpc})^3$, implying a $\sim$50\% chance of forming a low-mass cluster by $z= 0$ \citep[see Table 4 in][]{2013ApJ...779..127C}. As its effective radius is 2 cMpc — about half the typical size expected for a lower-mass cluster progenitor — this region might also evolve into a lower-mass structure, such as a group.

If we adopt the LAE field density reported by \cite{Herrera_2025} or the one calculated based on the luminosity function derived from the SC4K sample of \cite{Sobral_2018}, the regions S1, S2, S3, and S4 exhibit more pronounced overdensities. In Figure \ref{FIG:density_map}, the overdensity peak inside region S1 reaches $\delta \sim 9$ and $10.5$ when adopting \cite{Herrera_2025} and \cite{Sobral_2018}, respectively. However, the spatial extent set by the $2\sigma$ contour remains unchanged, and the average overdensity values in these regions show no relevant changes. Therefore, the overall evolutionary scenario previously discussed is not significantly altered. Accordingly, region S1 remains consistent with a moderate-mass cluster progenitor, and regions S2 and S4 with group progenitors. The main difference is that region S3 shows an increase in the average overdensity to $\bar{\delta} \sim 5.2$ and $6$ when adopting the field densities from \cite{Herrera_2025} and \cite{Sobral_2018}, respectively. This implies a higher probability of S3 evolving into a moderate-mass cluster, similarly to S1.

\subsection{Discovery of a LAE-traced extension of Taralay protocluster?}

Previously, \cite{Smol_2017} and \cite{Jimenez-Andrade_2023} reported an overdensity around J1000+0234 based on the photometric redshifts (\citealt{Laigle_2016} and \citealt{COSMOS2020}, respectively) of the sources in the COSMOS field. \cite{Jimenez-Andrade_2023} identified this overdensity as being centered on approximately 500 ckpc away from J1000+0234, extending over a radius of 5 cMpc. They found it to be consistent with a protocluster core, though with a low probability ($< 20\%$) of evolving into a low-mass cluster by $z=0$. They also pointed that, at this redshift, the submm source would be consistent with the evolutionary pathway in which high-redshift ($z>$ 3) SMGs evolve into massive elliptical galaxies \citep{Toft_2014, Stach_2021} that are typically found at the centers of galaxy clusters. In this work, we detect the Ly$\alpha$ blob \citep{Capak_2008,Jimenez-Andrade_2023}, one of the LAEs serendipitously identified by \cite{Jimenez-Andrade_2023} and three other nearby LAE candidates within a radius of approximately 1.2 arcmin ($\sim$ 2.7 cMpc) from J1000+0234. We also conclude that the local overdensity (S2) harboring the source J1000+0234 and its associated Ly$\alpha$ blob shows a low probability of evolving into a cluster, given its modest extent, and is more likely to evolve into a lower-mass structure, such as a galaxy group. However, when considering the global overdensity — comprising regions S1, S2, and S3 — within the wide area enclosed by the 2$\sigma$ contour described in Section \ref{sec:mass from simulations}, the system matches the scales of a high-mass cluster progenitor, as described by \cite{2015MNRAS.452.2528M}. This finding is in line with other works, reporting LABs inhabiting group halos at high redshift \citep[e.g., ][]{Matsuda_2004,Matsuda_2006, Prescott_2012, Guaita_2022} and infalling protogroups that may be accreted by a more massive protocluster, as pointed by \cite{Bădescu_2017}.

The overdense regions identified in this work are located at the same redshift of Taralay protocluster, $z\sim$ 4.5. This protocluster was first identified by \citet{Lemaux_2018} as PCl J1001+0220, based on spectroscopic data from the VIMOS 
Ultra-Deep Survey (VUDS). \citet{Staab_24_Taralay} refined its extent and internal structure, characterizing it as a massive structure in the redshift range 4.48 $<$ $z<$ 4.64 with an estimated mass of 1.7 $\times$ 10$^{15}$ M$_\odot$, which makes it exceptionally massive for these redshifts. At the same redshift and merely $\sim$ 10 cMpc away from Taralay protocluster, 
the LAE peak overdensity identified in this work may be tracing the dynamic state expected for a structure in the midst of being built. This supports a large-scale structure formation scenario in which more recently assembled halos hosting young, star-forming galaxies traced by LAEs fall into a more massive ancient halo.

\section{Conclusions}

We carry out a deep narrowband imaging of a circular 0.16 deg² section of the COSMOS field around the submm source J1000+0234. We test whether this SMG traces an overdense region at $z\simeq$ 4.54.
A total of 80 LAE candidates are identified at the J1000+0234 redshift ($z= 4.54 \pm 0.03$), 7 of which had been previously spectroscopically identified within this redshift range. The distribution of the sources forms 4 overdense regions within the surveyed area. The 2 densest regions, S1 and S3, span areas of 27 $\times$ 20 cMpc$^2$ and 25 $\times$ 15 cMpc$^2$, respectively. S1 exhibits an average overdensity of $\bar\delta$ = 3 within a volume of (25 cMpc)$^3$, while S3 shows an average overdensity of $\bar\delta$ = 3.8 within a volume of (15.7 cMpc)$^3$. The remaining two regions, S2 and S4, are less overdense, covering a radius of approximately 7 cMpc (S2) and an area of 15 $\times$ 10 cMpc$^2$ (S4). These regions show mean overdensities of $\bar\delta$ = 2.9 and $\bar\delta$ = 2, respectively, within a volume of (15 cMpc)³. Considering the dimensions described by \cite{2015MNRAS.452.2528M}, the extent of the region formed by S1, S2, and S3 — enclosed by a 2$\sigma$ overdensity contour — corresponds to the scale of a high-mass cluster progenitor, potentially reaching M $>$ 10$^{15}$ M$_\odot$ by $z= 0$. Alternatively, following \cite{2013ApJ...779..127C}, we find that the regions S1 and S3 are consistent with moderate- and low-mass cluster progenitors, respectively, and regions S2 and S4 are more consistent with progenitors of lower-mass structures, such as groups. The most overdense region, S1, is located at the same redshift, only $\sim$10 cMpc away from Taralay protocluster. These results suggest that we are witnessing the process of large-scale structure formation, in which recently assembled halos traced by LAEs are falling into the more massive halo of the Taralay protocluster. 
We conclude that the submm source located in S2, approximately 10 cMpc away from the major overdensity peak, thus traces a moderately overdense and potentially infalling protocluster.

Spectroscopic confirmation of the LAE candidates and the study of their properties within the overdense structure identified in this work will allow us to infer potential environmental effects acting on the formation of this large-scale structure. Additionally, NB surveys covering the redshift range of Taralay protocluster will allow us to probe the LAE population within this already massive protocluster, enabling an exploration of the role of halos hosting young star-forming galaxies, such as LAEs, and those hosting more mature, continuum-bright galaxies within the framework of hierarchical structure formation.

\begin{acknowledgments}
We thank the anonymous referee for the useful comments.
This research was financed in part by the Coordenação de Aperfeiçoamento de Pessoal de Nível Superior - Brasil (CAPES) - Finance Code 88882.425630/2019-01. KMD thanks the support of the Serrapilheira Institute (grant Serra-1709-17357) as well as that of the Brazilian National Research Council (CNPq grant 308584/2022-8) and of the Rio de Janeiro Research Foundation (FAPERJ grant E-32/200.952/2022), Brasil.
\end{acknowledgments}

\vspace{5mm}





\clearpage

\appendix
\renewcommand{\thefigure}{\thesection.\arabic{figure}} 
\section{Growth curve for the NB filter}
\label{Apendix:Growth curve for the NB filter}

\setcounter{figure}{0} 

\centering
\begin{figure}[h]
    \centering
    \includegraphics[scale=0.45]{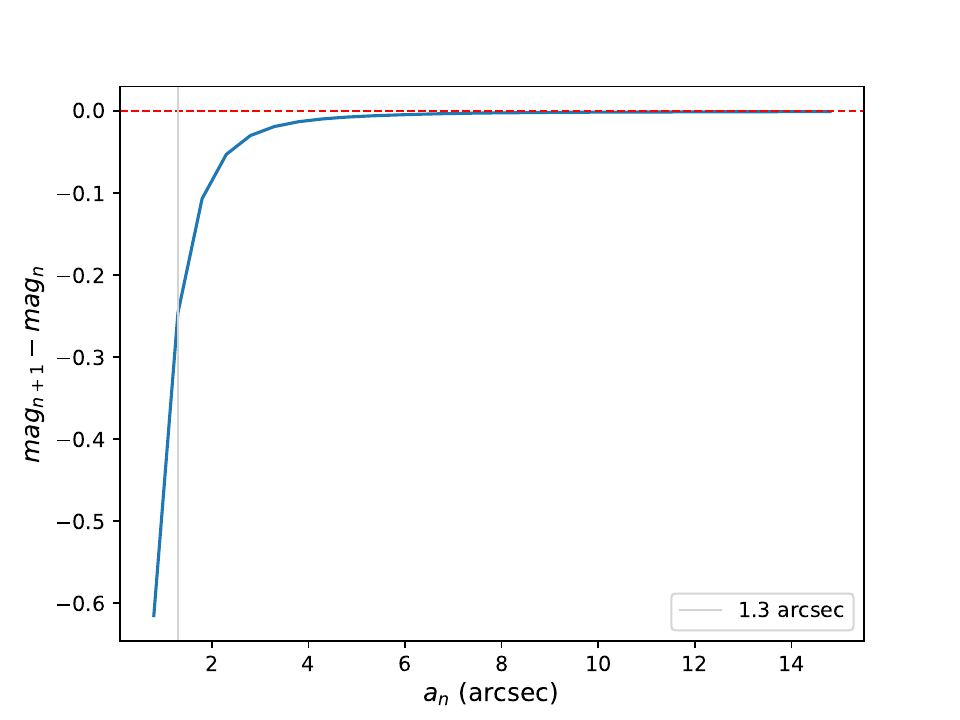}
    \caption{Growth Curve depicting the mean difference in instrumental magnitudes obtained with consecutive photometric apertures, differing by 0.5'' in diameter, $a_{n+1} = a_n + 0.5$'', as a function of photometric aperture. The highest S/N occurs at aperture 1.3'' (gray vertical line).}
    \label{FIG:APER_correction}
\end{figure}

\section{\texorpdfstring{Function \textit{\lowercase{$f$}(L)}}{Function f(L)}}

\label{Apendix:convolution}

\setcounter{figure}{0} 

\begin{figure}[h]
\centering

\includegraphics[scale=0.2499]{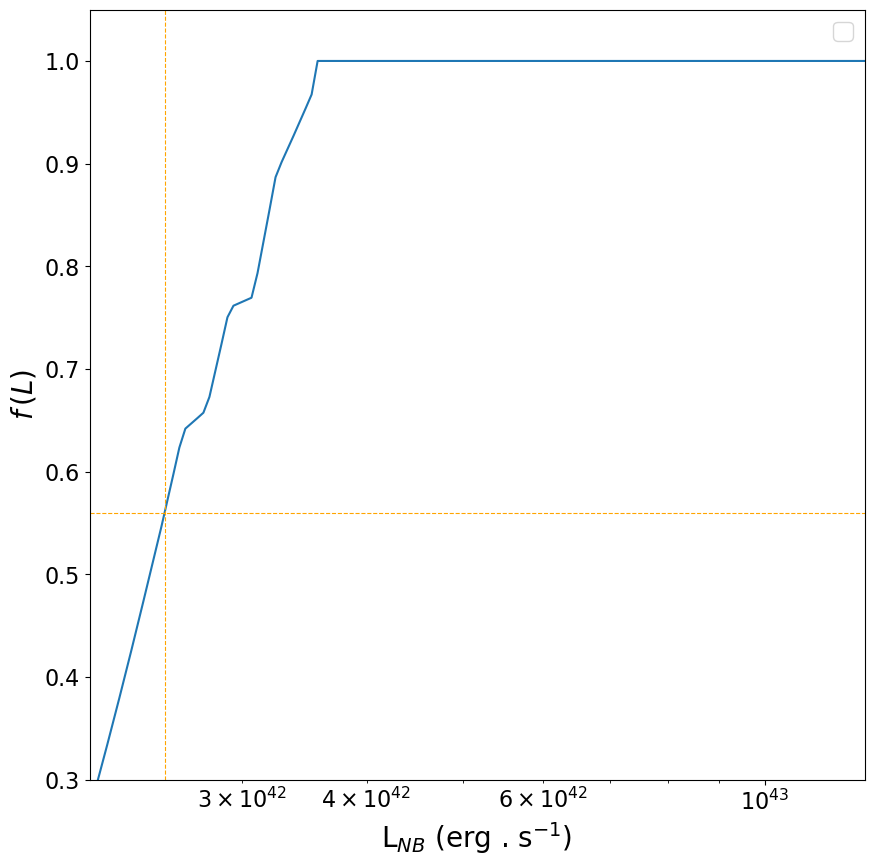}

 \caption{At a luminosity of $10^{42.37}$ erg s$^{-1}$,  corresponding to the magnitude cut of mag$_{NB}$ = 25.6, the function $f(L)$ has a value of 0.6. It increases with the NB filter's completeness, reaching 0.95 at a luminosity of $10^{42.57}$ erg s$^{-1}$, which corresponds to mag$_{NB}$ = 25.25. For higher luminosities, $f(L)$ equals 1.}
\label{Fig:function_f}
\end{figure}

\section{Effective radius}
\label{appendix:effective radius}

\begin{justify}

The effective radius $R_e$ of a protocluster as defined in \cite{2013ApJ...779..127C} is the square root of the second moment of the member halo positions, weighted by their masses. It is given by the equation:

$$R_e = \sqrt{\frac{1}{M} \sum m_i (x_i - x_c)^2}$$
where \(M\) is the total mass of the protocluster in bound halos, \(m_i\) is the mass of each halo, \(x_i\) is the position of each halo, and \(x_c\) is the center of mass of the protocluster. In the Lambda Cold Dark Matter ($\Lambda$CDM) model of the universe, galaxies are thought to reside within dark matter halos. We use the locations of galaxies to infer the likely positions of the dark matter halos in which these galaxies reside, and use the stellar mass provided by the COSMOS2020 catalog as an approximation of the halo mass.
\end{justify}

\vspace{0.5cm}

\section{Cutouts - LAE candidates}
\label{appendix:cutouts}

\setcounter{figure}{0} 

\begin{figure*}[h] 
    \centering
    \includegraphics[scale=0.35]{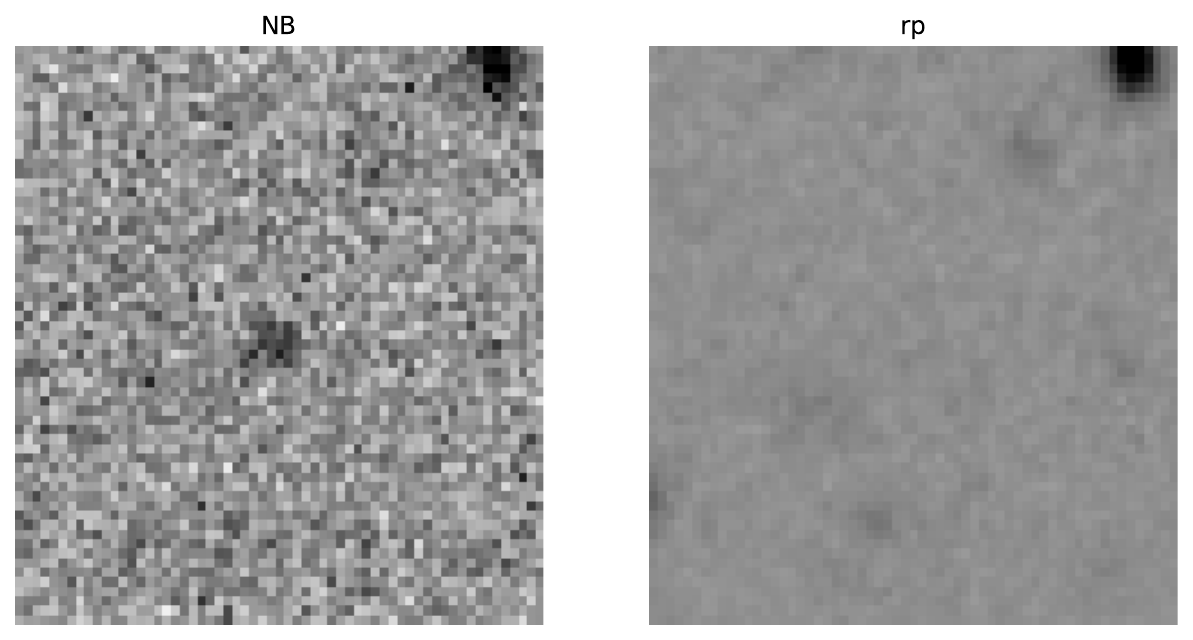}
    \hspace{1.5cm}\includegraphics[scale=0.35]{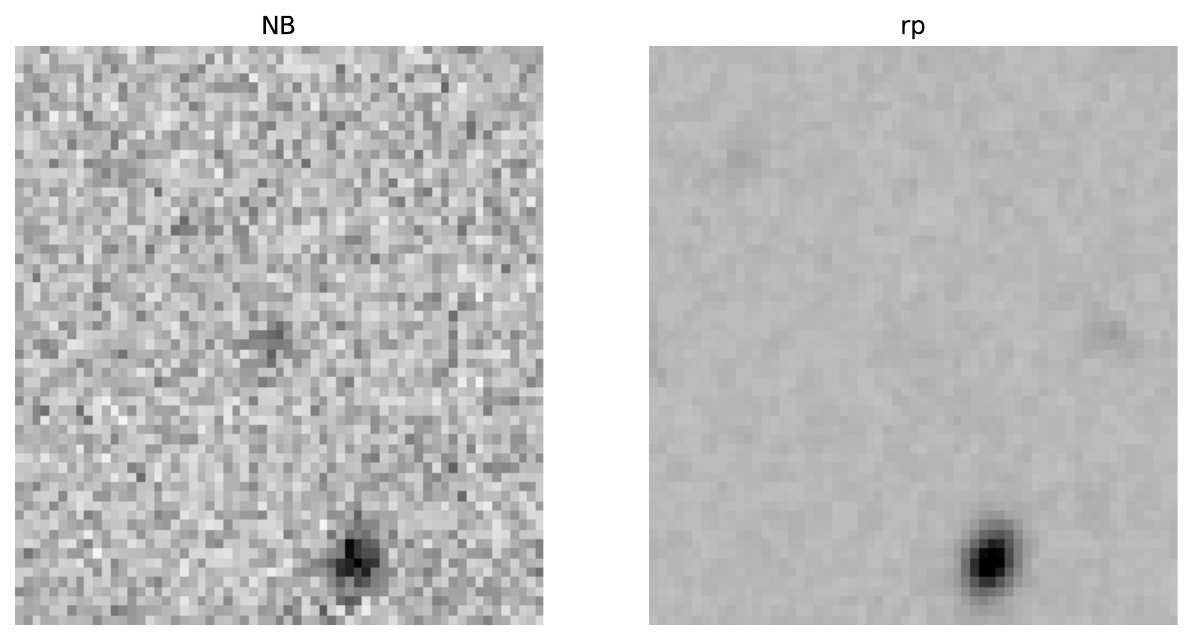}
    \includegraphics[scale=0.35]{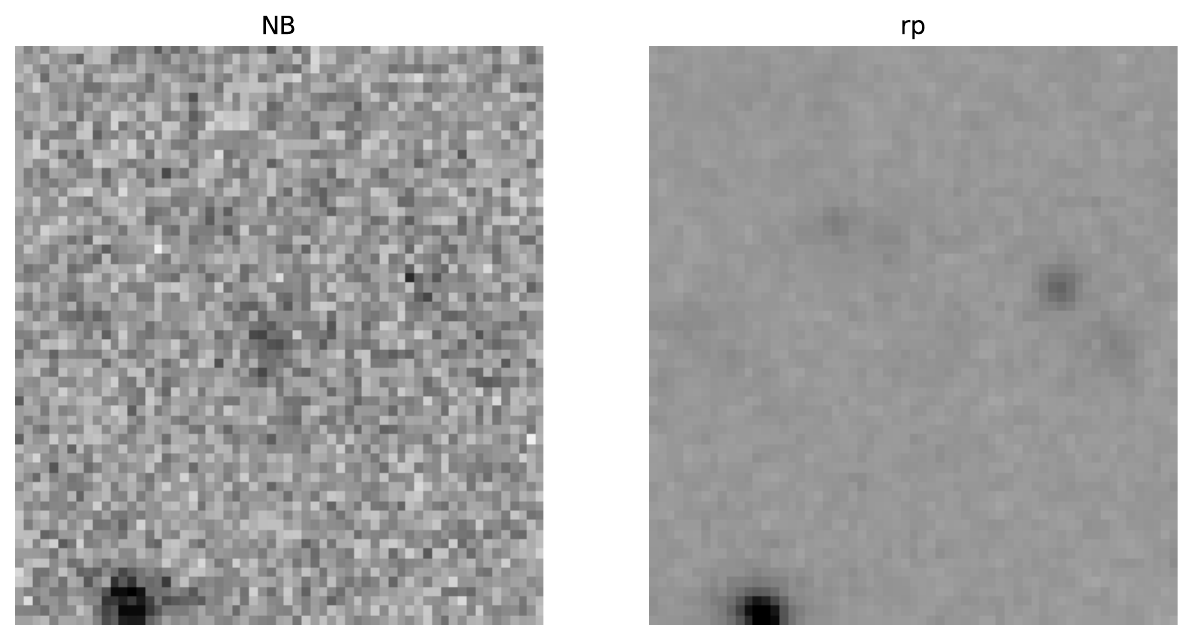}
    \hspace{1.5cm}\includegraphics[scale=0.35]{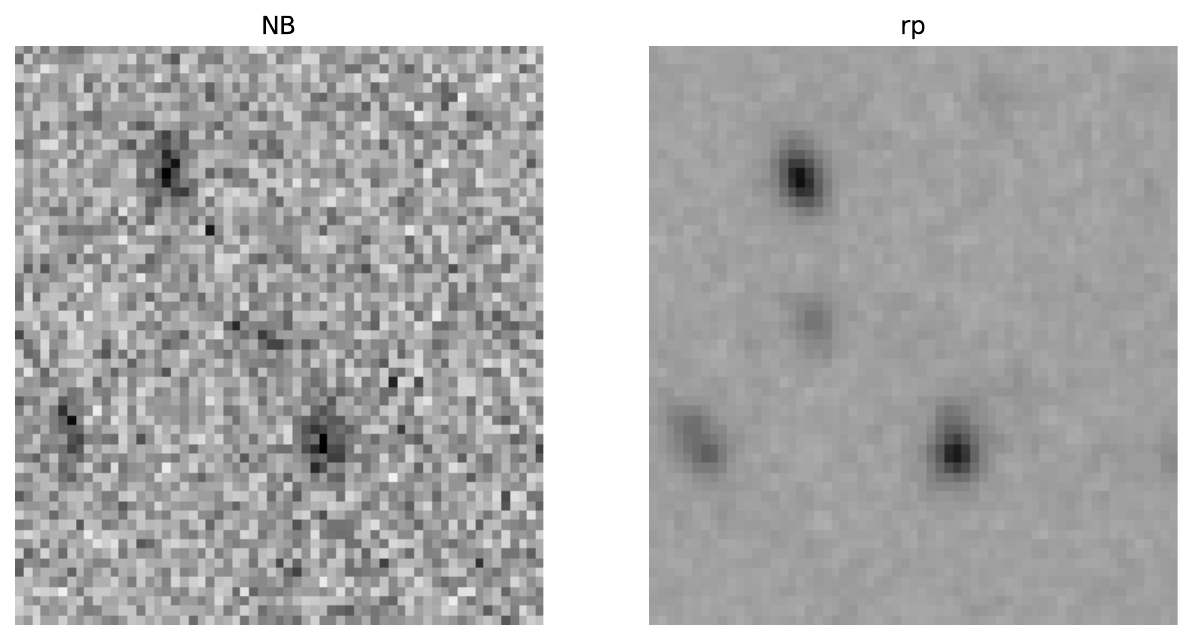}

    \includegraphics[scale=0.35]{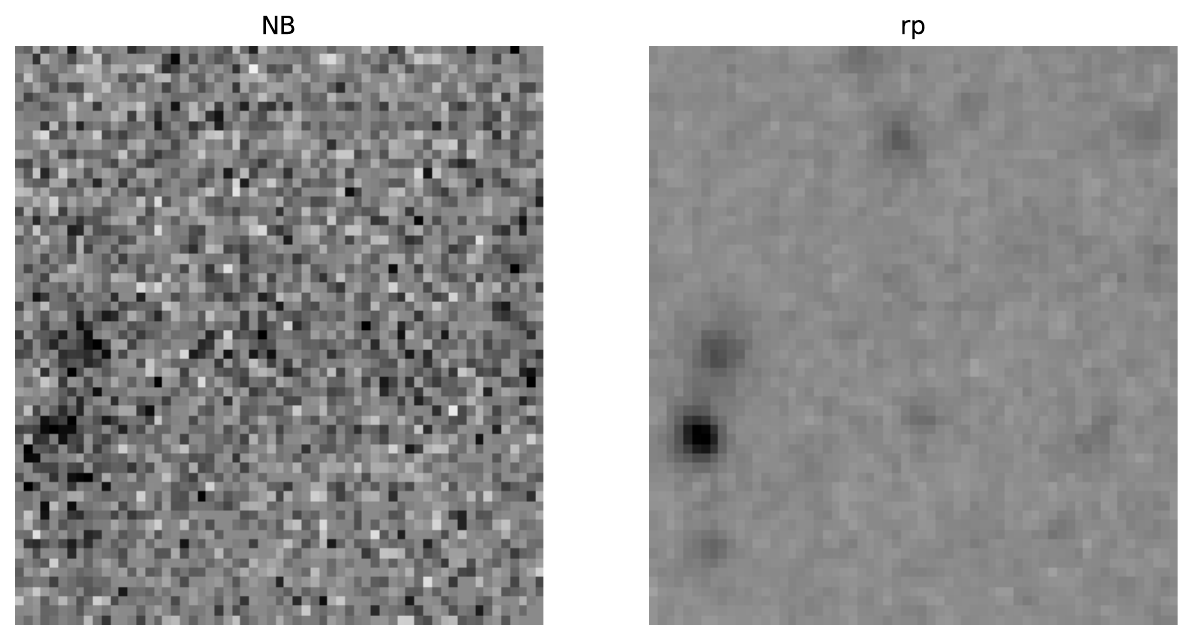}
    \hspace{1.5cm}\includegraphics[scale=0.35]{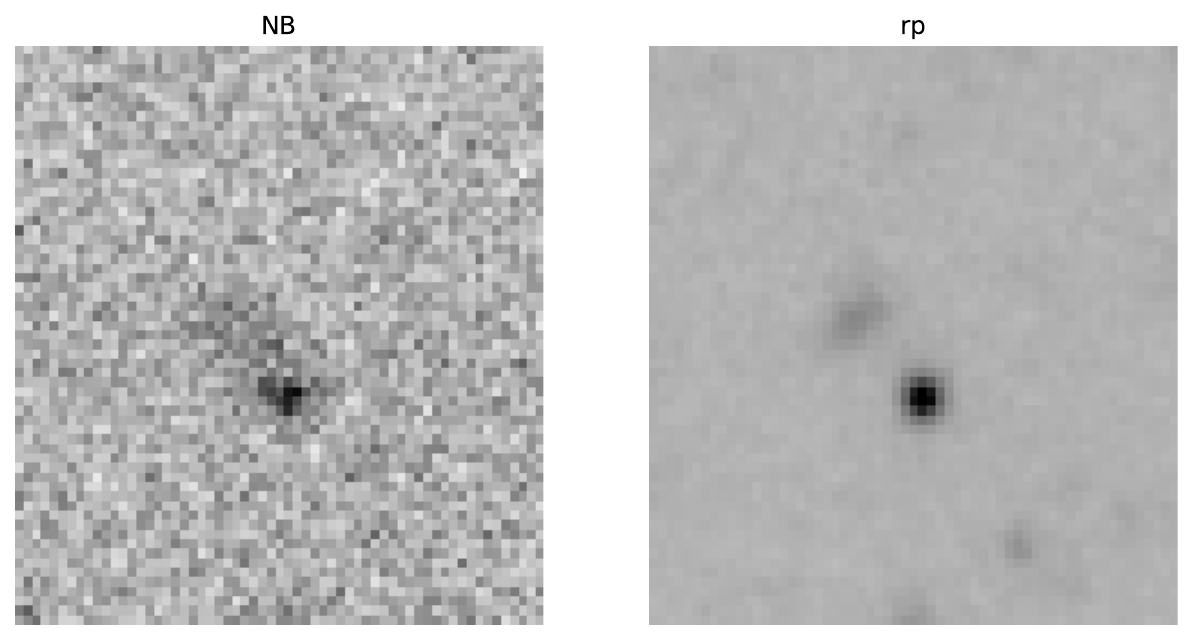}

   \includegraphics[scale=0.35]{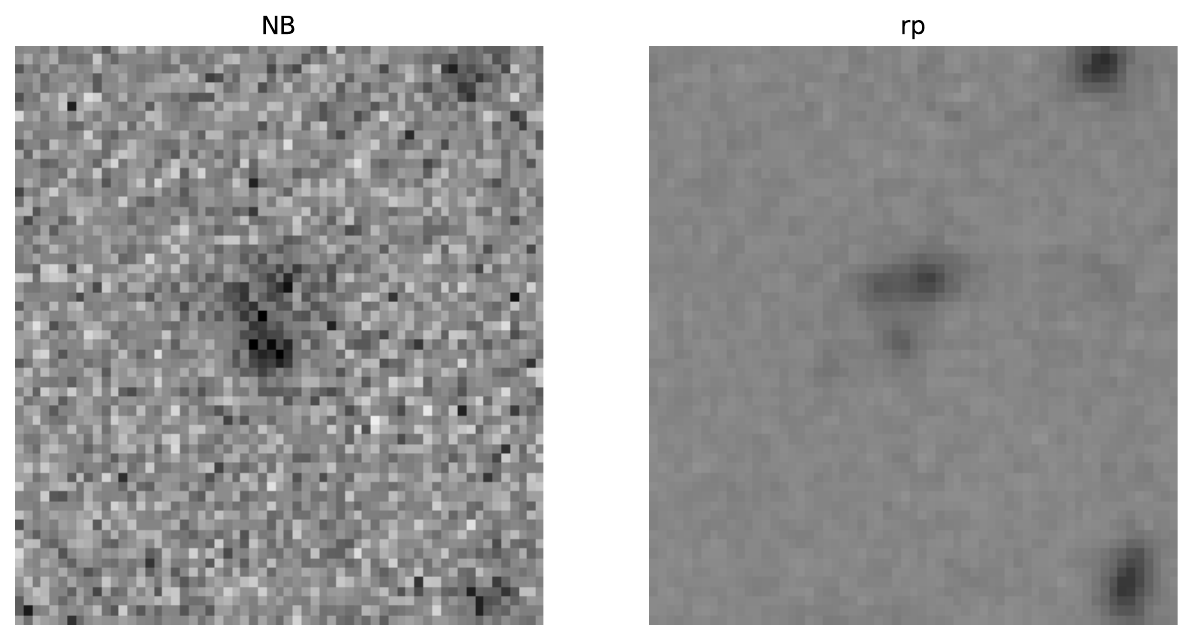}
    \hspace{1.5cm}\includegraphics[scale=0.35]{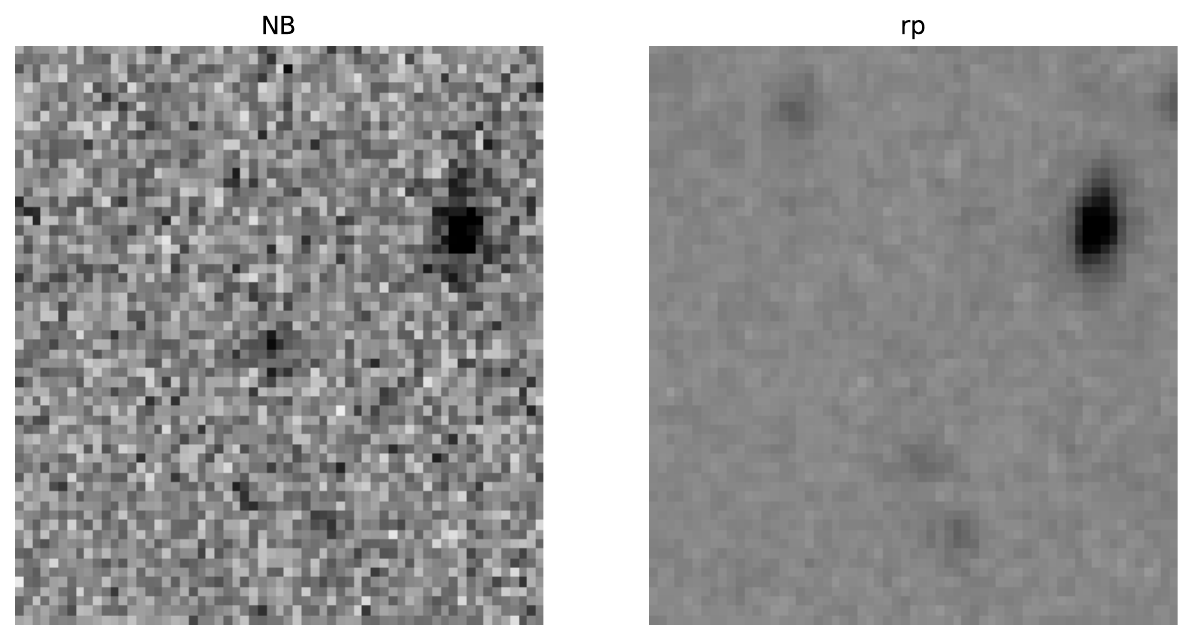}

    \caption{Cutouts of LAE candidates not included in the COSMOS2020 catalog. Each source is displayed side-by-side in NB and r+, respectively, with matched flux scaling for visual comparison. Each cutout is 12" × 12" in size. (Continued on the next page).
}
    \label{FIG:cutouts}
\end{figure*}



\setcounter{figure}{0} 

\begin{figure*}[p]
    \centering
    \includegraphics[scale=0.35]{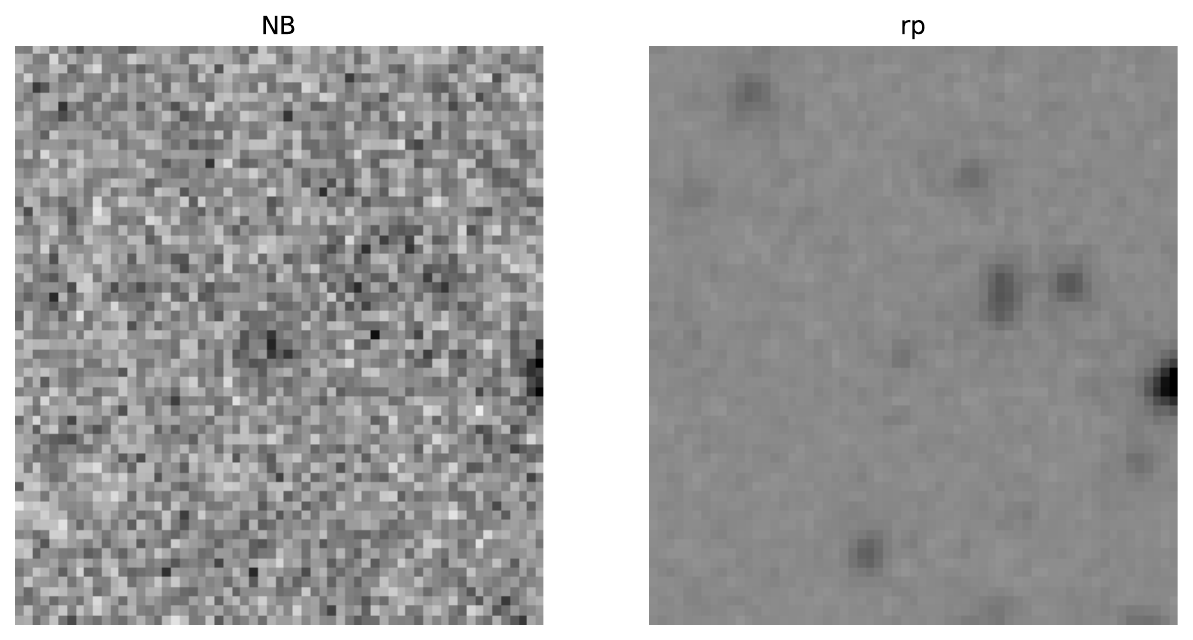}
    \hspace{1.5cm}\includegraphics[scale=0.35]{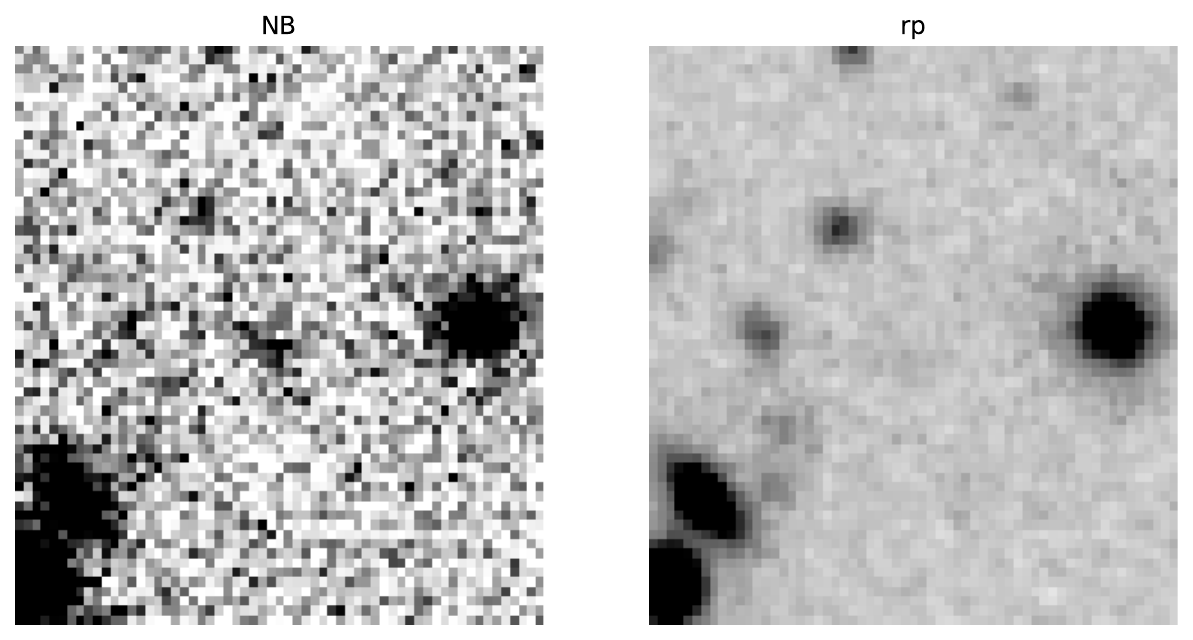}
    \includegraphics[scale=0.35]{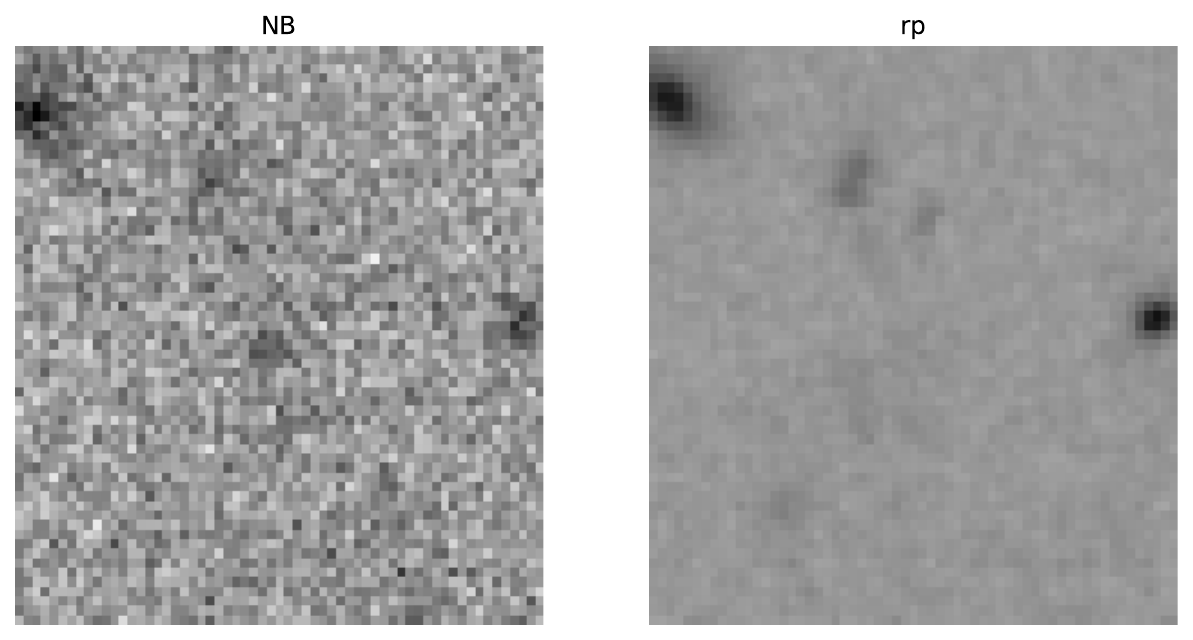}
    \hspace{1.5cm}\includegraphics[scale=0.35]{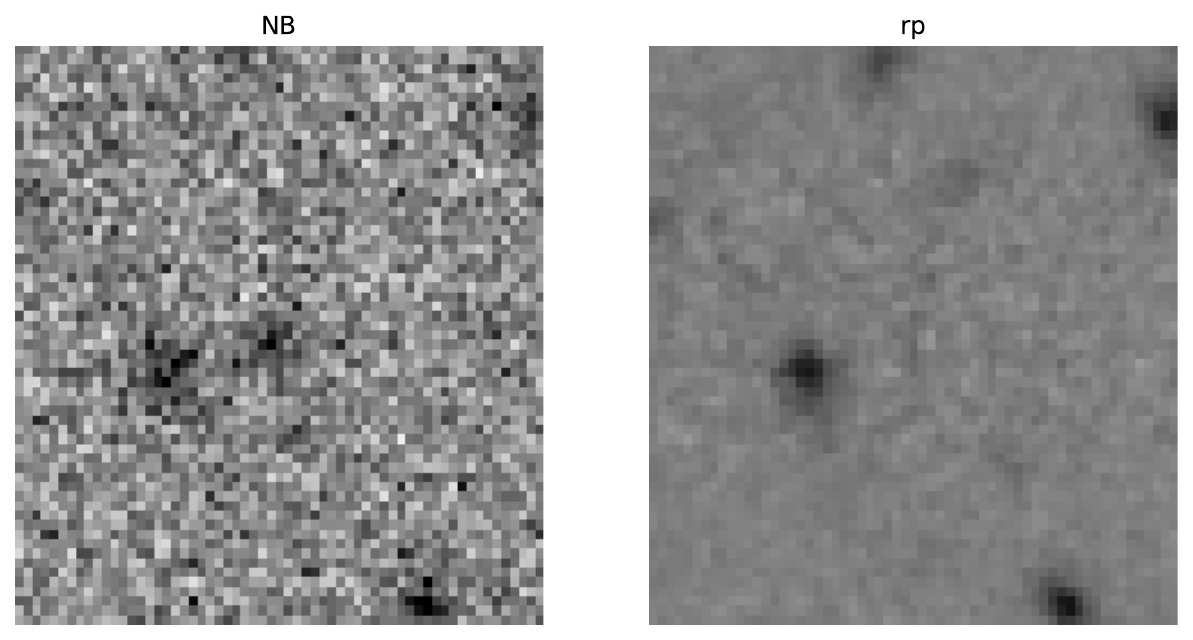}
    \includegraphics[scale=0.35]{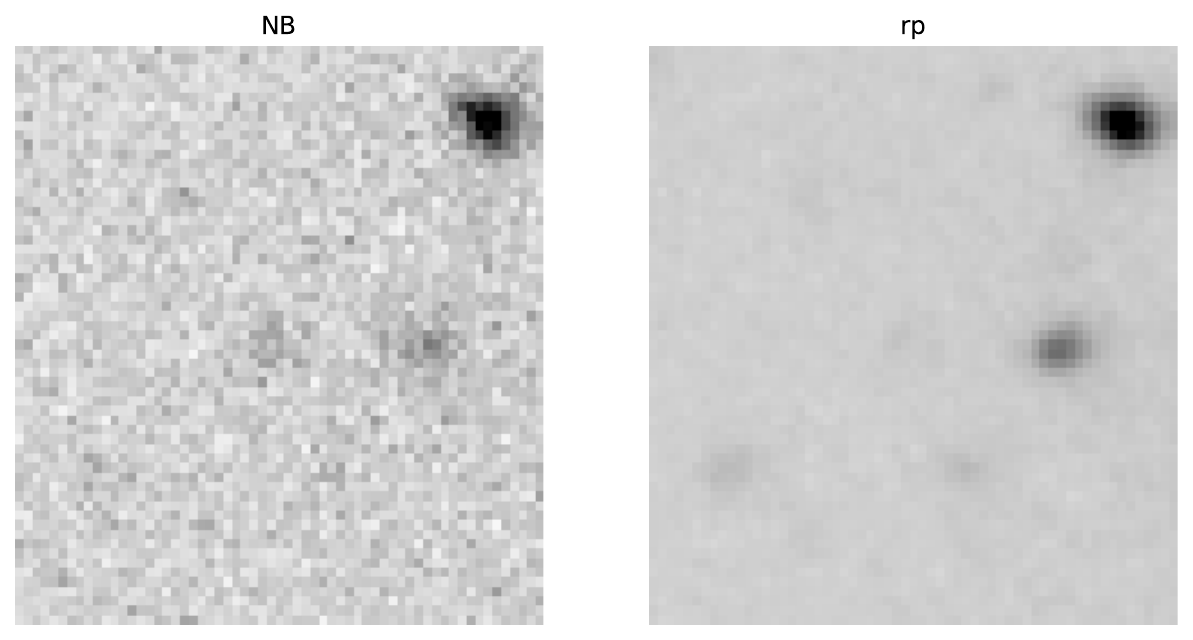}
    \hspace{1.5cm}\includegraphics[scale=0.35]{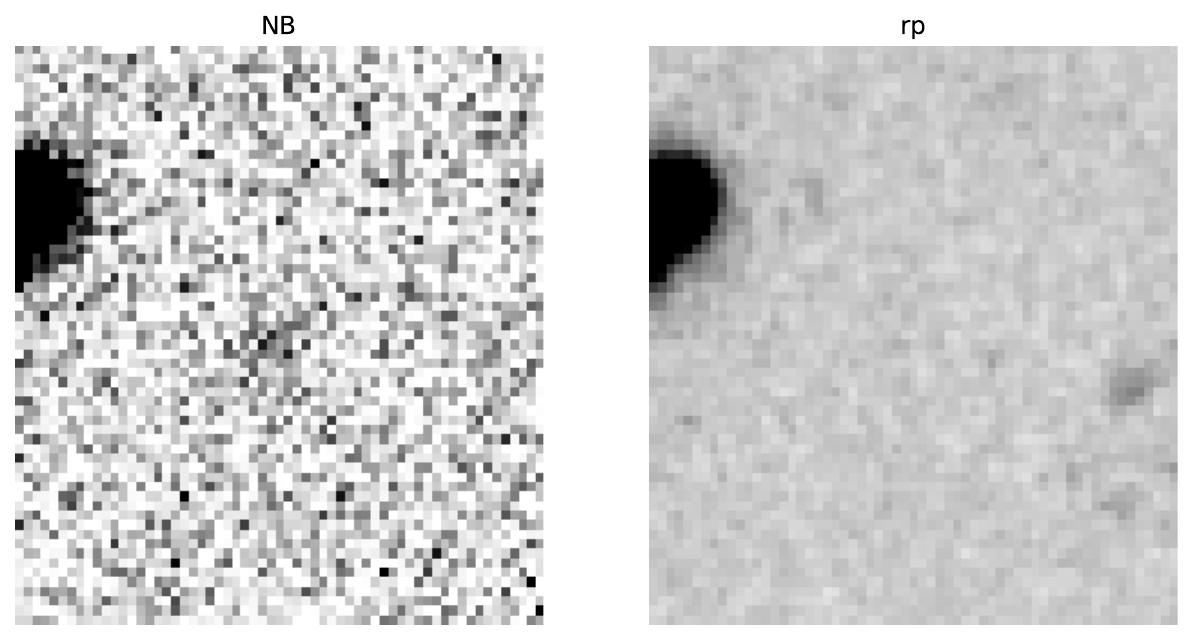}

    \includegraphics[scale=0.35]{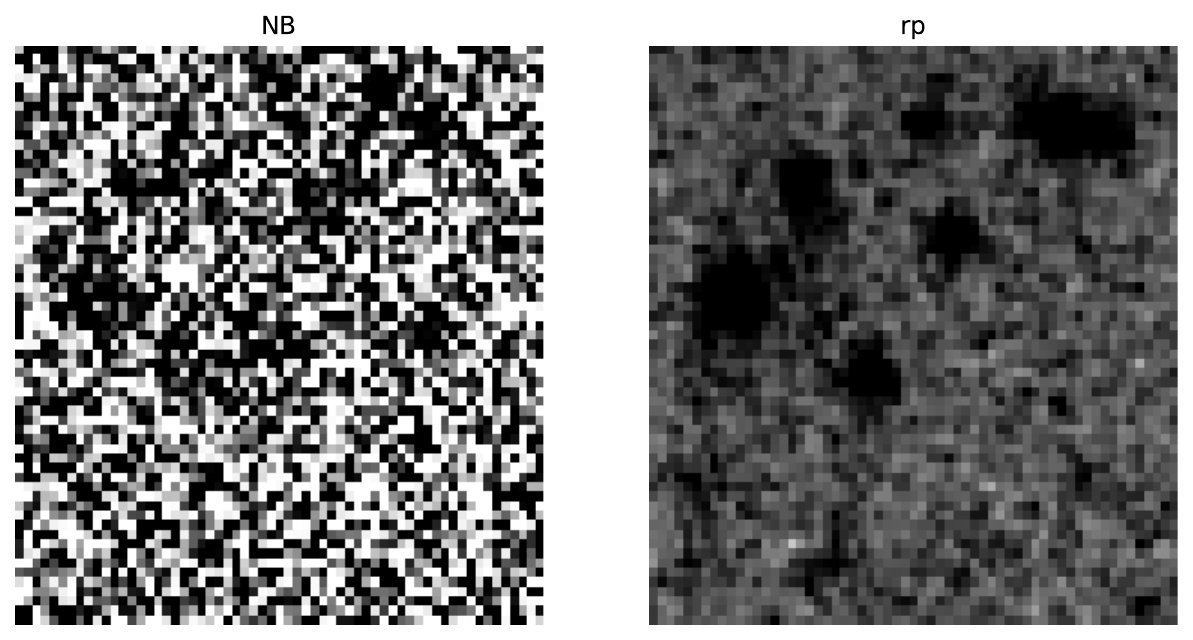}
    \hspace{1.5cm}\includegraphics[scale=0.35]{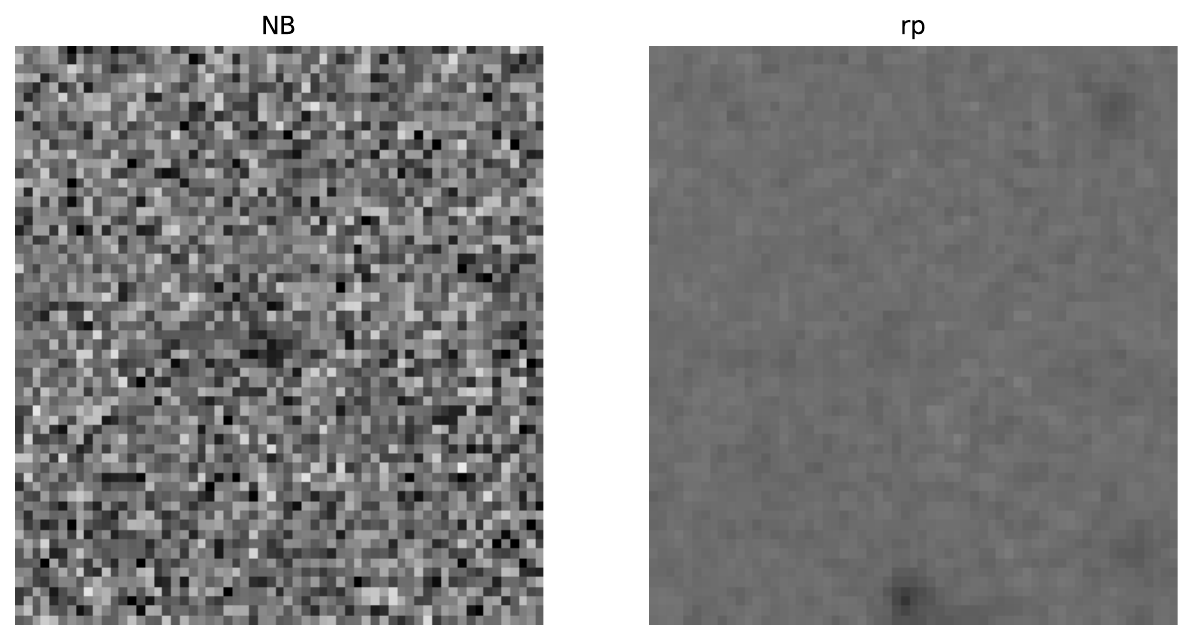}

    \includegraphics[scale=0.35]{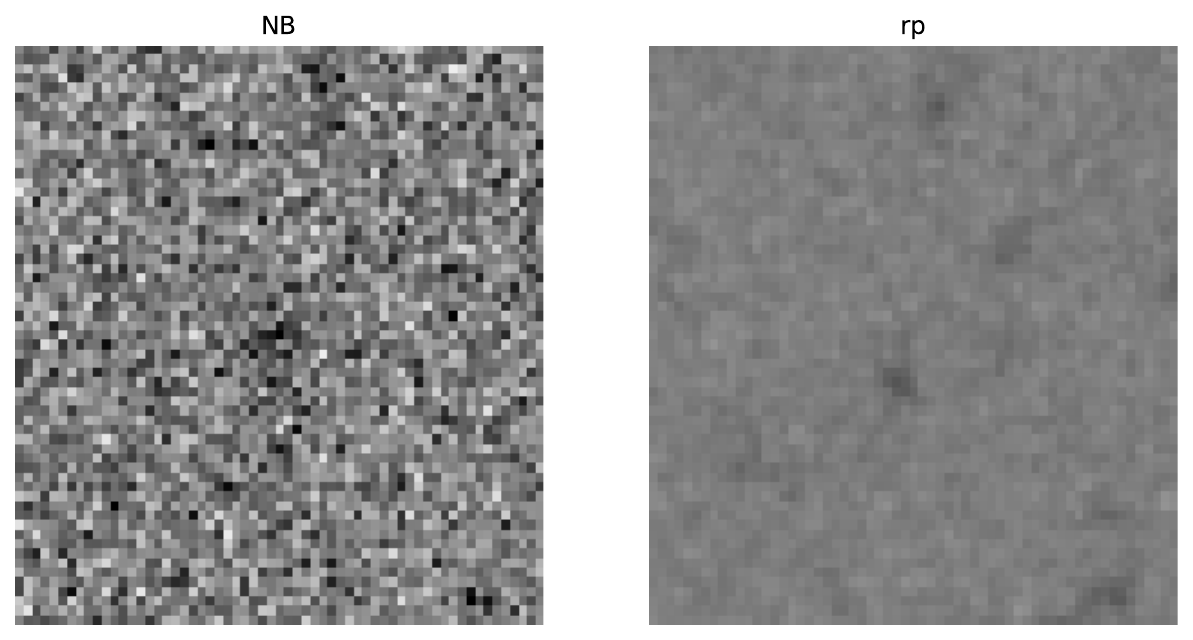}
    \hspace{1.5cm}\includegraphics[scale=0.35]{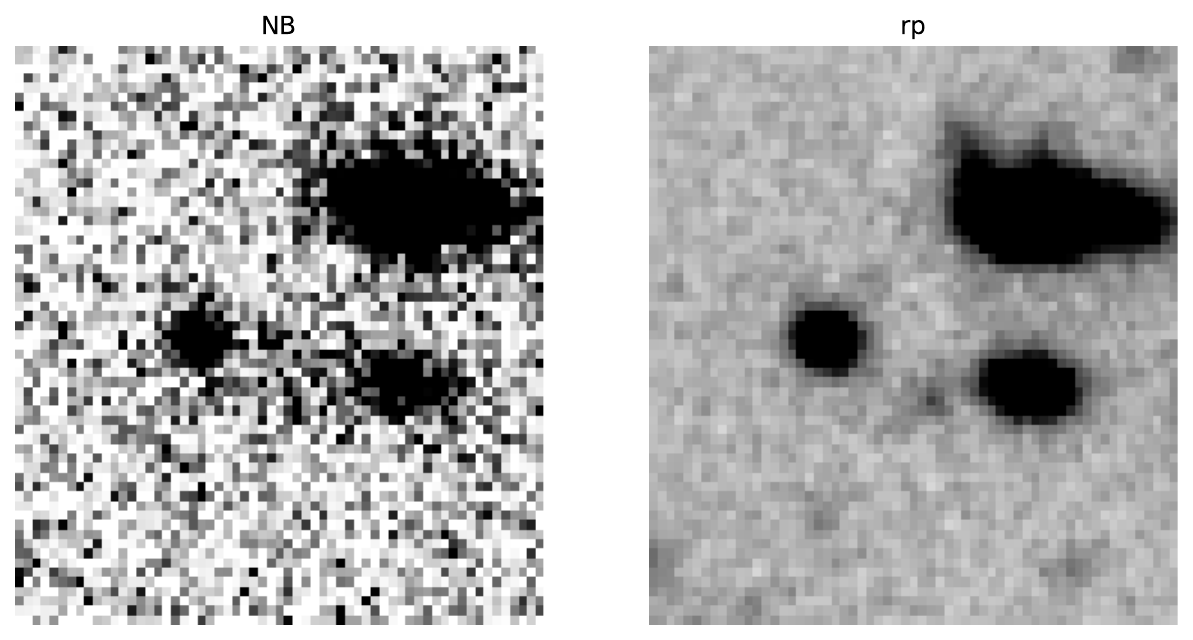}

    \caption{(Continued).\textcolor{white}{bbbbbbbbbbbbbbbbbbbbbbbbbbbbbbbbbbbbbbbbbbbbbbbbbbbbbbbbbbbbbb} } 
\end{figure*}

\newpage

\setcounter{figure}{0} 

\begin{figure*}[ht]
\centering

    \includegraphics[scale=0.35]{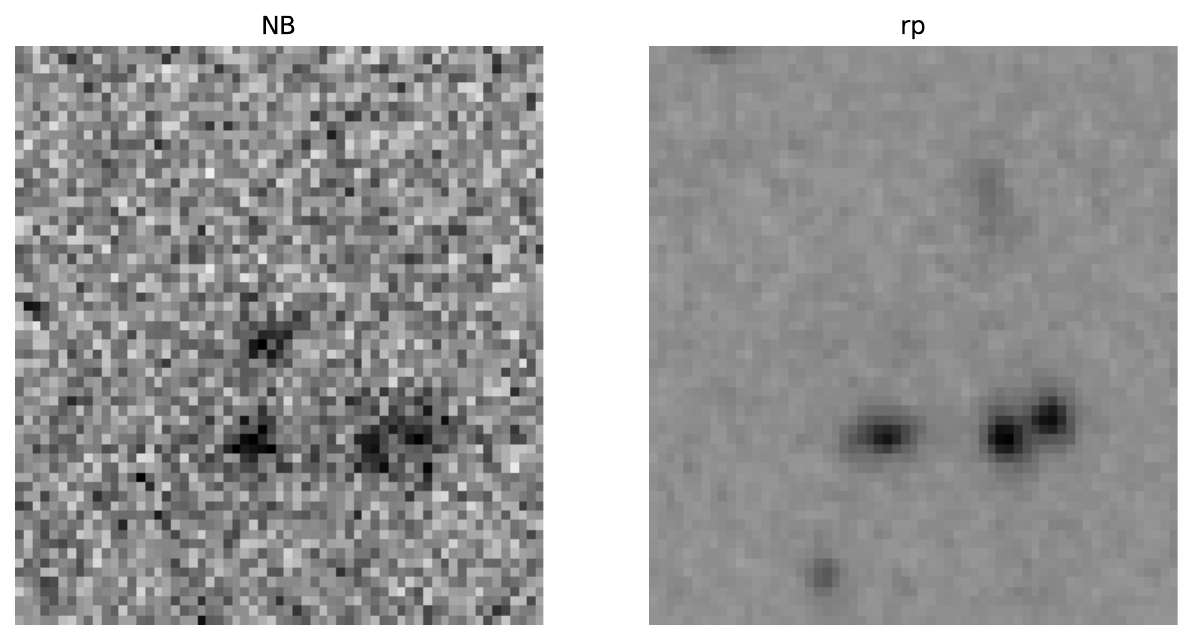}
    \hspace{1.5cm}\includegraphics[scale=0.35]{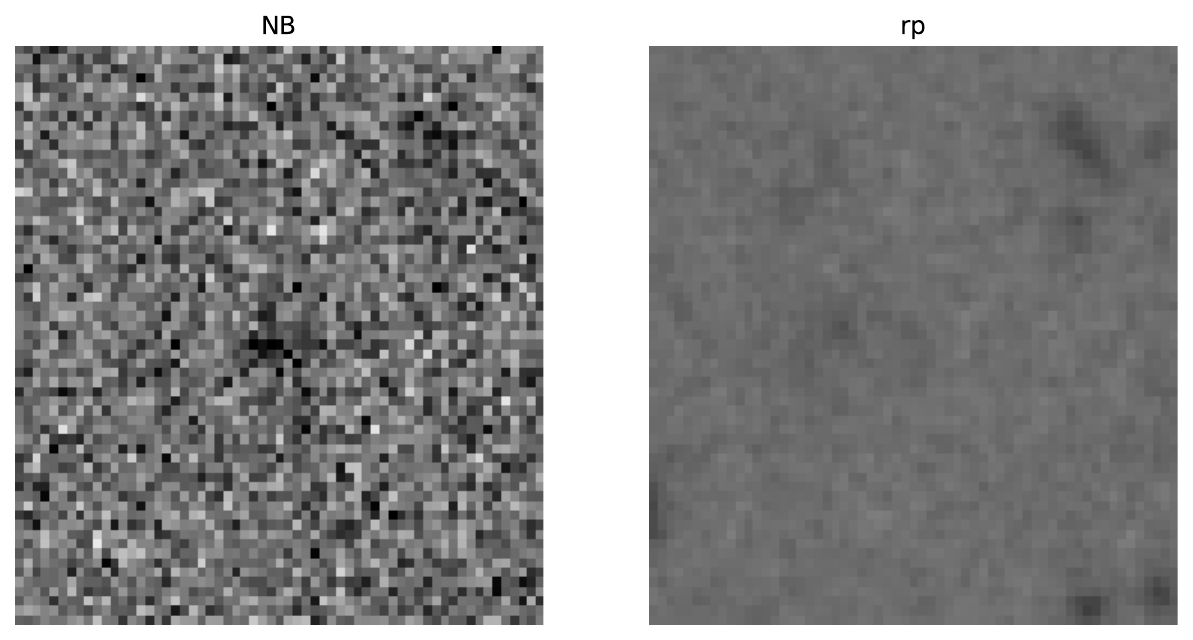}
  
    \caption{(Continued).\textcolor{white}{bbbbbbbbbbbbbbbbbbbbbbbbbbbbbbbbbbbbbbbbbbbbbbbbbbbbbbbbbbbbb}} 
\end{figure*}





\bibliographystyle{aasjournal}

\end{document}